\documentclass[11pt,numbers,sort&compress]{article}

\usepackage{latexsym,amsmath,amssymb,theorem,epsfig} 
\usepackage{natbib}
\usepackage{graphicx}

\DeclareFontFamily{OT1}{rsfs10}{}
\DeclareFontShape{OT1}{rsfs10}{m}{n}{ <-> rsfs10 }{}
\DeclareMathAlphabet{\mathscript}{OT1}{rsfs10}{m}{n}

\newcommand{\be}{\begin{equation}}
\newcommand{\beq}{\begin{equation}}
\newcommand{\ee}{\end{equation}}
\newcommand{\eeq}{\end{equation}}
\newcommand{\nn}{\nonumber}
\newcommand{\bea}{\begin{eqnarray}}
\newcommand{\eea}{\end{eqnarray}}
\newcommand{\ba}{\begin{array}}
\newcommand{\ea}{\end{array}}
\newcommand{\eref}[1]{(\ref{#1})}

\def\n{\nu}

\def\cA{{\cal A}}

\def\cN{{\cal N}}

\def\cC{{\cal C}}

\def\cL{{\cal L}}
\def\cV{{\cal V}}

\newcommand{\comment}[1]{}
\newcommand{\mycaption}[1]{\caption{{\sf \small #1}}}

\newcommand{\setall}{\setcounter{equation}{0}\setcounter{theorem}{0}}

\newcommand{\Tot}{\mathop{{\rm Tot}}}
\newcommand{\Cone}{\mathop{{\rm Cone}}}
\newcommand{\isom}{\cong}
\newcommand{\from}{\leftarrow}
\newcommand{\IP}{\mathbb{P}}

\newcommand{\IZ}{\mathbb{Z}}

\usepackage[all]{xy}
\newcommand{\rk}{\mathop{{\rm rk}}}
\newcommand{\coker}{\mathop{{\rm coker}}}

\newcommand{\im}{\mathop{{\rm im}}}
\newcommand{\cO}{{\cal O}}

\begin{document}

\begin{titlepage}
  \begin{flushright}
    arXiv:0904.2186
  \end{flushright}
  \vspace*{\stretch{1}}
  \begin{center}
     \huge {\bf Yukawa Couplings in Heterotic Compactification}
  \end{center}
\vspace{0.1cm}
  \begin{center}
    \begin{minipage}{\textwidth}
      \begin{center}
        Lara B. Anderson${}^{1,5}$,
        James Gray${}^2$,
        Dan Grayson${}^3$,
        Yang-Hui He${}^{2,4}$,
        Andr\'e Lukas${}^2$
       \end{center}
    \end{minipage}
  \end{center}
  \vspace*{1mm}
  \begin{center}
    \begin{minipage}{\textwidth}
{\small
      \begin{center}
        ${}^1$ Department of Physics and Astronomy,
        University of Pennsylvania,\\
        209 South 33rd Street, Philadelphia, PA 19104-6396, USA.\\[0.2cm]
        ${}^2$Rudolf Peierls Centre for Theoretical Physics, 
        University of Oxford,\\ 
        1 Keble Road, Oxford OX1 3NP, UK.\\[0.2cm]
        ${}^3$Department of Mathematics,
        University of Illinois at Urbana-Champaign,\\
        1409 W.~Green Street, Urbana IL 61801, USA.\\[0.2cm]
        ${}^4$Merton College, Oxford, OX1 4JD, UK.\\[0.2cm]
        ${}^5$ Institute for Advanced Study, School of Natural Sciences, \\
        Einstein Drive, Princeton, NJ, 08540, USA.\\[0.2cm]
      \end{center}
}
    \end{minipage}
  \end{center}
  \vspace*{\stretch{1}}
  \begin{abstract}
    \normalsize 
    We present a practical, algebraic method for efficiently
    calculating the Yukawa couplings of a large class of 
    heterotic compactifications on Calabi-Yau three-folds with
    non-standard embeddings. Our methodology covers all of,
    though is not restricted to, the recently  classified
    positive monads over favourable complete intersection
    Calabi-Yau three-folds. Since the algorithm is based on
    manipulating polynomials it can be easily implemented on a
    computer. This makes the automated investigation of Yukawa
    couplings for large classes of smooth heterotic
    compactifications a viable possibility.
  \end{abstract}
  \vspace*{\stretch{5}}
  \begin{minipage}{\textwidth}
    \underline{\hspace{5cm}}
    \\
    \footnotesize email: ${}^1$andlara@hep.physics.upenn.edu,
    ${}^2$j.gray1@physics.ox.ac.uk, hey@maths.ox.ac.uk,  \\
    \hspace{5cm} lukas@physics.ox.ac.uk, ${}^3$dan@math.uiuc.edu  \\
  \end{minipage}
\end{titlepage}


\tableofcontents

\section{Introduction}\setall

Some of the most important pieces of data defining a phenomenological
theory of particle physics are the Yukawa couplings. Since these
parameters determine particle masses and interactions, no theory's
phenomenology can be understood in even a rudimentary manner without
knowledge of their values. In string phenomenology, Yukawa couplings
are usually some of the first things one attempts to calculate once
the low energy particle spectrum of a model is known
\cite{Greene:1987xh,Candelas:1987se,Candelas:1987rx}.  However,
despite their importance, in many cases it is not known how to carry
out the calculations in practice.

For compactifications of heterotic string theory and M-theory on
Calabi-Yau three-folds, Yukawa couplings have been be calculated in
only a relatively small number of cases. Examples include orbifold
compactifications and heterotic models with ``standard embedding'',
that is, models where the gauge bundle is chosen to be the tangent
bundle of the Calabi-Yau manifold.  Aside from these, only a few
other, isolated, examples appear
\cite{Berglund:1990qh,Donagi:2004ia,Donagi:2006yf,McOrist:2008ji},
with some of these being closely related to the standard embedding.

In this paper, we considerably improve this situation by providing a
simple, easy to implement, algorithm for calculating the Yukawa
couplings in a large class of heterotic compactifications on smooth
Calabi-Yau spaces with ``non-standard embedding''. In such
compactifications the gauge fields are defined in terms of a general,
poly-stable holomorphic vector bundle \cite{Green:1987mn}. Our approach is
in the spirit of the recent papers~
\cite{Anderson:2007nc,Gabella:2008id}, where a systematic analysis of
such general heterotic compactifications by means of computational
algebraic geometry has been pursued.

The manifolds we will consider are the favourable complete
intersection Calabi-Yau (CICY) manifolds \cite{Candelas:1987kf}, which
comprise a set of 4515 three-folds. In addition, we consider vector
bundles defined over these manifolds that are built using the monad
construction \cite{monadbook}. For simplicity, we will present our
method for the case where certain bundle cohomology groups vanish, as
summarised in Table~\ref{conditions}, but it is likely that the basic
ideas can be extended to all stable bundles on CICY manifolds. These
conditions are automatically satisfied for positive monad bundles and
a large number of ``not too negative'' monad bundles. The class of
positive monad bundles has been recently studied in
references \cite{Anderson:2008uw,Anderson:2007nc,Anderson:2008ex,
  stabilitypaper}, and the methods described in the present paper
represent a further step towards a systematic analysis of their
phenomenological properties.

Our method calculates the Yukawa couplings that appear in the
superpotential of the four-dimensional theory. They are related to the
physical Yukawa couplings by a field rotation that brings the matter
field kinetic terms into canonical form. Unfortunately, the matter
field kinetic terms and, hence, the required field redefinitions, are
not explicitly known, so the physical Yukawa couplings cannot be
computed directly.  Numerical calculations~\cite{Numerical} may be the
only way to overcome this common limitation, and we have nothing more
to say about it in the present paper. In practice, this means that
only certain invariants of the Yukawa couplings, which are unchanged
under redefinition of the matter fields, can be regarded as
physical. For example, in cases where one can talk about a Yukawa
matrix, such as in a model with $SO (10)$ gauge group and a single
Higgs representation in ${\bf 10}$, the rank of this matrix is
physically meaningful.

\vspace{0.1cm}

Of the many different methods of constructing vector bundles, the monad construction is one that has been of consistent interest in the physics literature over the years (see, for example, references \cite{Distler:1987ee,Blumenhagen:2006ux,Blumenhagen:2005ga,Anderson:2008uw,Anderson:2007nc}). These constructions, which will be reviewed in section \ref{monadstuff}, lend themselves nicely to methods of computational
algebraic geometry. This feature, which allows us to systematically
study large classes of such bundles at a time, is one of main motivations for
focusing on monad bundles in this paper. We shall consider the cases of
$SU(n)$ bundles, where $n=3,4,5$, corresponding to the GUT visible
sector gauge groups $E_6$, $SO(10)$ and $SU(5)$, respectively.
We will give examples throughout our discussion, but in particular, in
section \ref{bigeg}, we give a detailed presentation of an $SO(10)$
model. We show how to engineer models with a single ${\bf 10}$ multiplet of $SO(10)$.
In addition, we show that the rank of the Yukawa matrix for the ${\bf 16}$ multiplets of a model engineered in this way is
one. As a result, these cases correspond to compactifications with one heavy family.

\vspace{0.1cm}

Before we delve into technicalities let us briefly outline the basic
method for computing Yukawa couplings, which is quite simple in
principle. For a heterotic compactification on a Calabi-Yau three-fold
$X$, the families can be identified as elements of the cohomology
group $H^1(X,V)$ of the gauge bundle $V$. Anti-families correspond to
$H^2(X,V)\simeq H^1(X,V^\star)$, but our focus will be on models
without anti-families, a property which is automatic for positive
monad bundles, as we will review. To be specific, let us discuss the
case of an $SU(3)$ bundle which leads to the low-energy gauge group
$E_6$ and families in ${\bf 27}$ representations. In this case, we are
interested in the ${\bf 27}^3$ Yukawa couplings, that is, we need to
understand the map $H^1(X,V)\times H^1(X,V)\times H^1(X,V)\rightarrow
H^3(X,\wedge^3 V)\simeq\mathbb{C}$ (the last equivalence holds because
$\wedge^3 V\simeq {\cal O}_X$ for an $SU(3)$ bundle $V$ and
$h^3(X,{\cal O}_X)=1$). It turns out, for monad bundles, that the
``family cohomology group'' $H^1(X,V)$ can be represented by a
quotient of polynomial spaces, containing polynomials of certain,
well-defined, degrees. Likewise, we can represent the ``Yukawa
cohomology group'' $H^3(X,\wedge^3 V)$ by a quotient of polynomial
spaces which, of course, must be one-dimensional. Let $Q$ be a
representative of the single class in this quotient and $P_I$, where
$I,J,K,\ldots =1,\dots ,h^1(X,V)$, a polynomial basis for the
families. Then the Yukawa couplings $\lambda_{IJK}$ are obtained by
multiplying three family representatives. The result represents an
element in the one-dimensional Yukawa quotient space and must, hence,
be proportional to $Q$. The constant of proportionality is precisely
the desired Yukawa coupling, so $[P_IP_JP_K]=\lambda_{IJK}[Q]$, where
$[\cdot ]$ denotes the class in the quotient space. Hence, calculating
Yukawa couplings is reduced to a simple procedure of multiplying
polynomials and projecting the result onto the class representative
$Q$. For the cases of bundles with structure groups $SU(4)$ and
$SU(5)$ the procedure is analogous although slightly more complicated.

\vspace{0.1cm}

The plan of this paper is as follows. In the next section, we introduce
the general methodology available for computing Yukawa couplings in heterotic
compactifications and the arena in which we shall be working: 
positive monad bundles over the complete intersection Calabi-Yau
manifolds. In section $3$, we proceed to outline the procedure for
calculating Yukawa couplings in such
compactifications. We split the discussion into several subsections -
one for each of the possible visible sector gauge groups of
interest ($E_6$, $SO(10)$, and $SU(5)$). Section 4 contains a detailed discussion of a one-Higgs
$SO(10)$ model. In section 5 we end with conclusions and prospects. A technical result required in the bulk
of the text, as well as the proof that the polynomial-based procedure,
outlined in section 3, indeed reproduces the physical Yukawa couplings
are presented in the Appendix.

\section{Yukawa couplings in heterotic compactification}\label{s:yuk}
\setall After a brief review of heterotic compactifications and Yukawa
couplings, in \S\ref{s:coho} we will describe how the problem of
calculating Yukawa couplings can be rephrased in terms of bundle
cohomology groups. We will also hint at our method for calculating
these interactions based thereon, leaving the technical details of the
actual procedure to the following section. In addition, in
\S\ref{monadstuff}, we shall describe the basic geometrical setup of
our class of Calabi-Yau manifolds and bundles.

\vspace{0.1cm}

Let us start by considering how Yukawa couplings are usually described
in heterotic compactifications. The matter fields in Calabi-Yau
compactifications of heterotic string theory and M-theory descend from the
internal parts of the gauge fields and their superpartners. In the case
where we have a visible sector gauge bundle $V$ over the Calabi-Yau
threefold $X$ taking values in a subgroup $G$ of $E_8$, the low energy
observable gauge group, $H$, is given by the commutant of $G$ in
$E_8$. The matter fields arise in the decomposition of the adjoint of
$E_8$ under $G \times H$: ${\bf 248} = \sum_I (R^I_{G}, R^I_{H})$ with
$R_G^I$ and $R^I_H$ being representations of the groups $G$ and $H$
respectively, indexed by $I$. See Table \ref{t:break} for a complete list of the decompositions of the ${\bf 248}$ of $E_8$ and associated cohomologies for standard heterotic theories.

\begin{table}[t]
\begin{center}
\begin{tabular}{|l|l|l|}
\hline
$G\times H$ & Breaking Pattern:  
${\bf 248}\rightarrow $ 
& Particle Spectrum\\  
\hline\hline
{\small $\rm{SU}(3)\times E_{6}$} & {\small $({\bf 1},{\bf
  78})\oplus ({\bf 3},{\bf 27})\oplus 
(\overline{\bf 3},\overline{\bf 27})\oplus ({\bf 8},{\bf 1})$ }
&
$
\ba{rcl}
n_{27}&=&h^{1}(V)\\ 
n_{\overline{27}}&=&h^{1}(V^\star)=h^{2}(V)\\
n_{1}&=&h^{1}(V\otimes V^\star)
\ea
$
\\  \hline
{\small $\rm{SU}(4)\times\rm{SO}(10)$} &{\small $({\bf 1},{\bf
  45})\oplus ({\bf 4},{\bf 16}) 
\oplus (\overline{\bf 4},\overline{\bf 16})\oplus ({\bf 6},{\bf
  10})\oplus ({\bf 15},{\bf 1})$ } 
&
$
\ba{rcl}
n_{16}&=&h^{1}(V)\\
n_{\overline{16}}&=&h^{1}(V^\star)=h^2(V)\\
n_{10}&=&h^{1}(\wedge ^{2}V)\\
n_{1}&=&h^{1}(V\otimes V^\star)
\ea
$
\\  \hline
{\small $\rm{SU}(5)\times\rm{SU}(5)$} &{\small $({\bf 1},{\bf 24})\oplus
({\bf 5},{\bf 10})\oplus (\overline{\bf 5},\overline{\bf 10})\oplus
({\bf 10},\overline{\bf 5})\oplus 
(\overline{\bf 10},{\bf 5})\oplus ({\bf 24},{\bf 1})$}
&
$
\ba{rcl}
n_{10}&=&h^{1}(V)\\ 
n_{\overline{10}}&=&h^{1}(V^\star)=h^2(V)\\ 
n_{5}&=&h^{1}(\wedge^{2}V^\star)\\
n_{\overline{5}}&=&h^{1}(\wedge ^{2}V)\\
n_{1}&=&h^{1}(V\otimes V^\star)
\ea
$
\\ \hline
\end{tabular}
\mycaption{{\label{t:break} A vector bundle $V$ with structure group
    $G$ can break the $E_8$ gauge group of the heterotic string into a
    GUT group $H$. The low-energy representations are found from the
    branching of the ${\bf 248}$ adjoint of $E_8$ under $G\times H$
    and the low-energy spectrum is obtained by computing the indicated
    bundle cohomology groups.}}
\end{center}
\end{table}
The reduction ansatz for the holomorphic part of the gauge field $A$
in 10-dimensions is, to lowest order \cite{Lukas:1997fg},
\bea \label{mrA} A = \sum_I C_I^i u_I^a T_{a i} + A_{\textnormal{BG}}
\;.\eea Here $A_{\textnormal{BG}}$ is the background gauge field
vacuum expectation value satisfying the hermitian Yang-Mills
equations. The first term in \eqref{mrA} gives rise to the
four-dimensional matter fields $C_I^i$, where $I$ is an index running
over the terms in the decomposition of ${\bf 248}$ above, and $i$ runs
over the dimension of each representation $R^I_{H}$. The $u_I^a$ are
bundle-valued harmonic 1-forms on $X$, taking values in the associated
representation $R^I_{G}$ of the bundle structure group $V$. Finally,
the $T_{i a}$ are the relevant generators of the broken part of the
original $E_8$ gauge group, that is, those broken generators that are
not part of the bundle group $G$. The objects of interest in this
paper are the trilinear couplings between the low energy matter fields
$C_I^i$.

A simple expression for the superpotential Yukawa couplings has been well known for a some time
\cite{Green:1987mn}:
\bea
\label{physyukawa}
\lambda_{I J K} \propto \int_X u^a_I \wedge u^b_{J} \wedge u^c_{K}
\wedge \bar{\Omega} f_{abc} \ .  \eea We have used a ``proportional
to'' sign here to emphasise the fact that, without knowledge of the
K\"ahler potential, we can not meaningfully make statements about the
overall normalization. In Eq.~\eqref{physyukawa}, the holomorphic
$(3,0)$ form has been denoted by $\Omega$ and the $f_{abc}$ are
constants descending from the structure constants of $E_8$, designed to
make the above expression invariant under the bundle group $G$.  This
is the form for these couplings in the low energy theory as given to
us by direct dimensional reduction. Naively, the evaluation of
\eqref{physyukawa} is computationally awkward. On a given Calabi-Yau
manifold, one would have to find explicit expressions for all of the
forms involved and then integrate over the manifold. For $(2,1)$
matter fields in standard embedding models this has been explicitly
carried out in references \cite{Candelas:1987se,Hubsch:1992nu}. To repeat
such an explicit calculation for non-standard embedding models would
be technically very challenging and we will instead pursue a
different, more algebraic approach.

\subsection{Rephrasing in terms of cohomologies}
\label{s:coho}
The formula \eqref{physyukawa} has many appealing properties
\cite{Green:1987mn}. In particular, it is quasi-topological\footnote{
  Indeed, for standard embedding models, the Yukawa couplings for
  $(1,1)$ matter fields are topological and are given by the triple
  intersection numbers of the Calabi-Yau manifold.}. It depends only
on the cohomology class of the 1-forms $u^a_I$ and not upon the actual
representative form within that chosen class. Indeed, taking $u^a_I
\to u^a_I + D \epsilon^a_I $, for example, one sees that the
change to \eqref{physyukawa}, \bea \label{cohonly} \int_X D
\epsilon^a_I \wedge u^b_J \wedge u^c_K \wedge \bar{\Omega} f_{abc} \;,
\eea vanishes upon integration by parts since both the 1-forms $u^b$
and the holomorphic 3-form $\bar{\Omega}$ are $D$ closed. Given this
observation, one can regard the matter fields as being represented in
the formula \eqref{physyukawa} by cohomology classes, and not just
their harmonic representatives. This suggests that a simple
description of Yukawa couplings in terms of topological quantities
exists.

To pursue this idea, we begin by rewriting the formula for the Yukawa
couplings in the case where the bundle structure group $G$ is
$SU(3)$. We can then calculate four dimensional couplings between
three ${\bf 27}$ multiplets of $E_6$. The relevant structure constants
in this case are $f_{abc}=\epsilon_{abc}$ and, hence, the combination
$u^a_I \wedge u^b_J \wedge u^c_K \epsilon_{abc}$ is an $SU(3)$
invariant harmonic 3-form. Up to an overall constant multiple there
is, of course, only one such form on a Calabi-Yau $3$-fold, namely the
$(3,0)$ form $\Omega$. Thus we have that, \bea
\label{mrmap} u^a_I \wedge u^b_J \wedge u^c_K \epsilon_{abc} = K_{IJK}
\Omega \;,
\eea
where $K_{IJK}$ are complex numbers. From
\bea
 \lambda_{IJK} \propto \int_X u^a_I \wedge
u^b_J \wedge u^c_K \wedge \bar{\Omega} \epsilon_{abc} = K_{IJK} \int_X \Omega
\wedge \bar{\Omega}
\eea
we see these numbers are proportional to the desired Yukawa couplings.

Referring to Table \ref{t:break} once more, we see that the families in the ${\bf 27}$ representation of $E_6$
can be identified with the cohomology group $H^1(X,V)$. Therefore,
equation \eqref{mrmap} defines a map that takes
three of our bundle-valued 1-forms to a harmonic
3-form valued in the trivial bundle.  Now, for an $SU(n)$ bundle $V$ we
have that $\wedge^n V \cong \cO_X$, where $\cO_X$ is the trivial line-bundle on $X$.
Thus, \eqref{mrmap} defines a map of the form
\bea 
\label{cohyuk} 
H^1(X, V) \times H^1(X, V) \times H^1(X, V) \to H^3(X,\wedge^3 V)\cong H^3(X,\cO_X)\cong\mathbb{C}\; ,
\eea 
where the last equivalence follows from the fact that $h^3(X,\cO_X)=1$.

The main point of this paper is that, for a large class of
compactifications, when the above cohomologies are represented by
certain polynomial equivalence classes, there is a mathematically
natural proposal for what the map implicit in Eq.~\eqref{cohyuk}
is. It is essentially the unique possibility and simply involves
polynomial multiplication of cohomology representatives. In the next
section, we present this proposal in detail and show that the results
to which it gives rise have all of the properties one would expect.
The rigorous proof that our method for calculating Yukawa couplings
does indeed reproduce the physical formula \eqref{physyukawa} is
somewhat technical and is thus presented in Appendix \ref{ap:equal}.

\vspace{0.1cm}

A similar procedure can be applied to the case of structure group
$G=SU(4)$ and a visible gauge group $SO(10)$. For such models, we are
interested in Yukawa couplings of the type ${\bf 10}\, {\bf 16}\, {\bf
  16}$, between two families in ${\bf 16}$ representations and a Higgs
multiplet in a ${\bf 10}$ representation of $SO(10)$. Note that the
absence of anti-families in our models means there are no
$\overline{\bf 16}$ representations. From Table \ref{t:break}, it is
clear that families are still identified with the cohomology group
$H^1(X,V)$ while Higgs multiplets correspond to $H^1(X,\wedge^2 V)$.
The analogue of Eq.~\eqref{cohyuk} is then
\begin{equation}
\label{cohyuk4} 
H^1(X, V) \times H^1(X, V) \times H^1(X, \wedge^2 V) \to H^3(X,\wedge^4 V)\cong H^3(X,\cO_X)\cong\mathbb{C}~. 
\end{equation}
The appearance of the fourth wedge power, $\wedge^4 V$, means that one has to deal with polynomials of quite high degree in practical calculations. For this reason, it is useful to slightly reformulate the above mapping to
\bea 
\label{so10cohyuk} 
H^1(X, V) \times H^1(X, V) \to (H^1(X, \wedge^2 V))^* \cong H^2(X, \wedge^2 V)\;.  
\eea 
where the final equivalence follows from Serre duality \cite{hart} , $H^p(X, W) \simeq H^{3-p}(X, W^*)^*$, and the fact that $\wedge^2 V\cong\wedge^2 V^\star$ for $SU(4)$ bundles. Hence, instead of mapping two families and a Higgs multiplet into a one-dimensional space of high degree we combine two families to represent an element in the Higgs cohomology group. The relevant Yukawa couplings are then given by expressing the result in terms of a basis of Higgs multiplets. In this case, we only need to deal with second wedge powers of $V$ which, as we will see, implies lower polynomial degrees. 

\vspace{0.1cm}

Finally, the case where $G = SU(5)$ can be dealt with in either of the
two ways we have discussed so far. It turns out to be computationally
more efficient to follow the second approach.  From Table
\ref{t:break} we have three relevant multiplets, namely ${\bf 10}$
multiplets associated to $H^1(X,V)$, ${\bf 5}$ multiplets associated
to $H^1(X,\wedge^2V^\star )$ and $\overline{\bf 5}$ multiplets
associated to $H^1(X,\wedge^2 V)$ (and since we are considering models
without anti-families there are no $\overline{\bf 10}$ representations
present). This gives rise to two types of Yukawa couplings that are
schematically of the form ${\bf 10}\, {\bf 10}\, {\bf 5}$ and
$\overline{\bf 5}\,\overline{\bf 5}\, {\bf 10}$. The corresponding
maps in cohomology are \bea H^1(X,V) \times H^1(X, V) &\to& (H^1(X,
\wedge^2 V^\star))^* \cong H^2(X, \wedge^2 V) \label{cohyuk5}\\
H^1(X, \wedge^2 V) \times H^1(X, \wedge^2 V) &\to& (H^1(X, V))^* \cong
H^2(X, \wedge^4 V) \label{cohyuklast} \eea We now need to discuss how
the maps implied in \eqref{cohyuk}, \eqref{so10cohyuk},
\eqref{cohyuk5} and \eqref{cohyuklast} can actually be carried out
explicitly. As we will see, within our class of models provided by
CICY manifolds and monad bundles, the various cohomology groups can be
represented by quotient spaces of polynomials and the maps amount to
polynomial multiplication. To explore this in detail we now briefly
describe the technical arena we will be working in - that of positive
monad bundles over CICY manifolds - before we return to the problem of
calculating Yukawa couplings in \S\ref{s:calc}.
\subsection{The arena: positive monad bundles over CICYs}
\label{monadstuff}

In this paper, we will focus on heterotic compactifications involving
vector bundles built via the monad construction \cite{monadbook}. In
particular, we consider the class of positive monads\footnote{For
  reviews of this construction and some of its applications, see
  references ~\cite{monadbook,Anderson:2007nc, Blumenhagen:2006ux}.}
defined over favourable CICY manifolds \cite{Candelas:1987kf}. A
systematic analysis of the stability and spectrum of this class has
recently been completed in
\cite{Anderson:2007nc,Anderson:2008uw,stabilitypaper,Anderson:2008ex}.

To begin, we recall that complete intersection CICY manifolds are defined by the zero loci of $K$ polynomials $\{p_j\}_{j=1,\ldots ,K}$ in an ambient space  $\cA =\IP^{n_1} \times \ldots \times \IP^{n_m}$ given by a product of $m$  projective spaces with dimensions $n_r$. We denote the projective coordinates of each factor $\IP^{n_r}$ by $(x_0^{(r)},x_1^{(r)},\ldots,x_{n_r}^{(r)})$, its K\"ahler form by $J_r$, and the $k^{\rm th}$ power of the hyperplane bundle by $\cO_{\IP^{n_r}}(k)$. The manifold $X$ is called a {\it complete intersection} if the dimension of $X$ equals the dimension of $\cA$ minus the number of polynomials. To obtain three-folds $X$  in this way we then need $\sum_{r=1}^m n_r - K = 3$.

Each of the defining homogeneous polynomials $p_j$ can be characterised by its multi-degree ${\bf q}_j=(q_j^1,\ldots , q_j^m)$, where $q_j^r$ specifies the degree of $p_j$ in the coordinates ${\bf x}^{(r)}$ of the factor $\IP^{n_r}$ in $\cA$.  These polynomial degrees are conveniently encoded in a configuration matrix
\beq\label{cy-config}
\left[\ba{c|cccc}
\IP^{n_1} & q_{1}^{1} & q_{2}^{1} & \ldots & q_{K}^{1} \\
\IP^{n_2} & q_{1}^{2} & q_{2}^{2} & \ldots & q_{K}^{2} \\
\vdots & \vdots & \vdots & \ddots & \vdots \\
\IP^{n_m} & q_{1}^{m} & q_{2}^{m} & \ldots & q_{K}^{m} \\
\ea\right]_{m \times K}\; .
\eeq
Note that the $j^{\rm th}$ column of this matrix contains the multi-degree of the polynomial $p_j$. The Calabi-Yau condition, $c_1(TX)=0$, is equivalent to the conditions $\sum_{j=1}^Kq^r_j=n_r+1$. In terms of this data, the normal bundle $\cN$ of the CICY manifold $X$ in $\cA$ can be written as
\begin{equation}
 \cN = \bigoplus_{j=1}^K\cO_\cA ({\bf q}_j)\; . \label{normalbundle}
\end{equation}
Here and in the following we employ the short-hand notation $\cO_\cA
({\bf k})=\cO_{\IP^{n_1}}(k^1)\otimes\dots\otimes\cO_{\IP^{n_r}}(k^r)$
for line bundles on the ambient space $\cA$. In the notation given
above, the famous quintic hypersurface in $\IP^4$ is denoted as ``$[4
| 5]$'' and its normal bundle is $\cN=\cO_{\mathbb{P}^4}(5)$.

CICY threefolds have been completely classified \cite{Candelas:1987kf}
and of the 7890 manifolds, $4515$ are favourable, that is, all of
their K\"ahler forms, $J$, descend from those of the ambient
projective space. This means that favourable CICY manifolds defined in
an ambient space with $m$ projective factors are characterized by
$h^{1,1}(TX)=m$. We will focus on these favourable CICY manifolds in
the following.

A monad bundle, $V$, is defined by the short exact sequence
\bea
\nn &&0 \to V \to B \stackrel{f}{\longrightarrow} C \to 0\ ,
\mbox{ where} \\
B &=& \bigoplus_{i=1}^{r_B} \cO_X({\bf b}_i) \ , \quad
C = \bigoplus_{j=1}^{r_C} \cO_X({\bf c}_j)
\label{monad}
\eea are sums of line bundles with ranks $r_B$ and $r_C$,
respectively\footnote{More generally, a monad bundle is defined as the
  middle homology of a sequence of the form $0\to A
  \stackrel{m_1}{\longrightarrow} B \to C\to 0$. This sequence is
  exact at $A$ and $C$, and $\textnormal{Im}(m_1)$ is a subbundle of
  $B$ \cite{monadbook}. In this paper we restrict ourselves, as is
  often done in the physics literature, to the case where
  $\textnormal{Im}(m_1)$~vanishes. We thus recover the description
  \eqref{monad}.}. From the exactness of \eref{monad}, it follows that
the bundle $V$ is given by \beq\label{kern} V=\ker(f) \ .  \eeq From
the above sequence, the rank, $n$, of $V$ is \beq n = \rk(V) = r_B -
r_C \ .  \eeq For the structure group to be $SU(n)$ rather than $U(n)$
we need the first Chern class of $V$ to vanish, hence
\begin{equation}
 c_1^r(V)=\sum_{i=1}^{r_B}b_i^r-\sum_{a=1}^{r_C}c_a^r=0\; .
\end{equation} 
The existence of sufficiently general maps $f$ is guaranteed by demanding that $c_j^r\geq b_i^s$ $\forall i,j,r,s$.  We can think of $f$ as a matrix $f_{ai}$ of polynomials with multi-degree ${\bf c}_a-{\bf b}_i$. Furthermore, from Eq.~\eref{kern}, the bundle moduli of $V$ can be identified as the coefficients parameterizing the possible maps $f$ (see \cite{Anderson:2008uw} for a discussion). The term ``positive'' refers to monad bundles satisfying $b^r_i>0$ and $c^r_j>0$ $\forall r,i,j$.

For the technical details of monad bundles, including the spectrum,
moduli and such properties as slope-stability, we refer the reader to
\cite{Anderson:2007nc,Anderson:2008uw,stabilitypaper,Anderson:2008ex}. Here
we will review one feature of positive monad bundles that will be of
use to us in the following sections:
\begin{quote}
{\em Positive monads do not give rise to anti-generations, that is, $H^2(X,V)=H^1(X,V^*)=0$.}
\end{quote}

To see this, we consider dual of the monad sequence \eref{monad}
\bea 0 \to C^* \to B^* \to V^* \to 0 \;, \eea 
which gives rise to a long exact sequence
\bea
\label{cohoV*} \ldots \to H^1(X, B^*) \to H^1(X, V^*) \to H^2(X, C^*) \to \ldots\;.  
\eea
Now, since $B$ and $C$ are sums of positive line bundles both $H^1(X, B^*)$ and $H^2(X, C^*)$ are zero from  Kodaira's vanishing theorem (see, for example, references \cite{hart,Hubsch:1992nu}) so that $H^1(X,V^\star)=0$ follows immediately. Hence, there are no anti-families.

With these preliminary definitions in hand we turn now to the calculation of Yukawa couplings.

\section{Calculating Yukawa couplings: general procedure}\label{s:calc}
\setall We shall consider in turn the three types of theories with
$E_6$, $SO(10)$ and $SU(5)$ low-energy groups, corresponding
respectively to the choices of an $SU(n)$ bundle structure group with
$n=3,4,5$. A concrete $SU(3)$ example will be presented in this
section but, in the interests of brevity, we postpone doing the same
for the more complicated $SO(10)$ case until the next section. We do
not give a detailed $SU(5)$ example in this paper because the
techniques are lengthy, while qualitatively the same as in the $SU(3)$
and $SU(4)$ cases.

While the idea of computing Yukawa couplings using polynomial methods
is based on the sheaf-module correspondence and should be quite
general and widely applicable, the specific realisation discussed in
this paper relies on a number of vanishing properties which we
summarise in Table~\ref{conditions}.
\begin{table}
\begin{center}
\begin{tabular}{|c|c|}\hline
Case &Cohomologies required to vanish \\
\hline \hline $E_6$ & $H^1(X, B)$,  $H^3(X, \wedge^3 B)$, $H^2(X,
\wedge^3 B)$ \\ &$H^1(X, \wedge^2 B \otimes C)$,  $H^2(X, \wedge^2 B
\otimes C)$, $H^1(X, B \otimes S^2 C)$ \\ \hline 
$SO(10)$ & $H^1(X,B)$,  $H^1(X, \wedge^2 B)$, $H^2(X, \wedge^2 B)$, $H^1(B \otimes C)$ \\ \hline 
$SU(5)$& $H^1(X, B)$, $H^1(X, \wedge^2 B)$, $H^1(X, \wedge^4 B)$ \\ &
$H^2(X, \wedge^4 B)$, $H^1(X, \wedge^3 B \otimes C)$ \\ & 
$H^2(X, \wedge^2 B)$, $H^1(X, B \otimes C)$\\ \hline 
\end{tabular}
\mycaption{\label{conditions} List of vanishing conditions on the sums of line bundles $B$ and $C$, defining the monad bundle, required for our calculation. All conditions are automatically satisfied for positive monads, due to the Kodaira vanishing theorem.}
\end{center}
\end{table}  
These conditions are all automatically satisfied for positive monad bundles $V$, that is, when the sums of line bundles $B$ and $C$ that enter the monad sequence~\eqref{monad} consist of positive line bundles only. 

Given these conditions, we would like to derive polynomial representations for certain bundle cohomology groups and maps between them.  It is useful to first discuss this problem for the main building blocks of the monad construction, line bundles.

\subsection{Polynomial representation of line bundle cohomology}
We begin with the simple case of a single projective space $\mathbb{P}^n$ with projective coordinates ${\bf x}=(x_0,\ldots ,x_n)$ and an associated graded ring $R=\mathbb{C}[{\bf x}]$. It is well-known that the sections, $H^0(\mathbb{P}^n,\cO_{\mathbb{P}^n}(k))$ of the line bundle $\cO_{\mathbb{P}^n}(k)$ can be identified with the degree $k$ polynomials in $R$. We denote the degree $k$ part of $R$ by $R_k$ and write $H^0(\mathbb{P}^n,\cO_{\mathbb{P}^n}(k))\cong R_k$.

The generalization to products of projective spaces, ${\cal A}=\mathbb{P}^{n_1}\times\dots\times\mathbb{P}^{n_m}$, is straightforward. We denote the projective coordinates of the $r^{\rm th}$ projective space by ${\bf x}^{(r)}$ and the associated multi-graded ring by
\begin{equation}
 R=\mathbb{C}[{\bf x}^{(1)},\ldots ,{\bf x}^{(m)}]\; . \label{Rring}
\end{equation}
Then the sections of the line bundle $\cO_\cA ({\bf k})$ can be identified with the multi-degree ${\bf k}$ polynomials in $R$, so
\begin{equation}
H^0(\cA ,\cO_\cA ({\bf k}))\cong R_{\bf k}\; . \label{HR}
\end{equation} 

In our actual applications, we are of course interested in line bundles $\cO_X ({\bf k})$ on the CICY manifold $X\subset\cA$. They can be related to their ambient space cousins via a Koszul resolution and this leads to a method of calculating their cohomology and, in particular, their sections. The details of this argument are given in Appendix \ref{ap:koszul} but the final result is rather simple. Consider the polynomial ring~\eqref{Rring}, associated to our ambient space $\cA$, and the ideal $\langle p_1,\ldots ,p_K\rangle\subset R$ generated by the defining polynomials $p_j$ of the CICY manifold $X$. Then we can form the coordinate ring
\begin{equation}
 A=\frac{R}{\langle p_1,\ldots ,p_K\rangle}
\end{equation}
of the CICY manifold $X$, which one can think of as the space of
polynomials on $X$. In terms of the coordinate ring, the sections of
the line bundle $\cO_X({\bf k})$ are given by
\begin{equation}
 H^0(X,\cO_X({\bf k}))\cong A_{\bf k}\; , \label{HA}
\end{equation}
where the $A_{\bf k}$ denotes the multi-degree ${\bf k}$ part of $A$. This relation requires certain vanishing conditions, as detailed in Appendix \ref{ap:koszul}, which are all automatically satisfied for positive line bundles. The result~\eqref{HA} is in close analogy to its ambient space counterpart~\eqref{HR}, so all that is required when dealing with line bundles on the CICY manifold $X$ is passing from the full polynomial ring to the coordinate ring of $X$. 

\subsection{$SU(3)$ vector bundles and $E_6$ GUTS} \label{e6}
We start by considering the case of $SU(3)$ bundles. From Table \ref{t:break}, the symmetry breaking pattern and decomposition of the matter field representations is 
\bea
E_8 &\supset& SU(3) \times E_6 \\
{\bf 248} &=& ({\bf 8},{\bf 1}) \oplus ({\bf 1},{\bf 78}) \oplus ({\bf 3},{\bf 27}) \oplus
(\overline{\bf 3},\overline{\bf 27})
\eea
The $({\bf 8},{\bf 1})$ term in this decomposition is associated with the cohomology
group $H^1(X, V \otimes V^*)$ that counts the dimension of the bundle moduli space.
Furthermore, when we are dealing with positive monads, as discussed above, anti-families in $\overline{\bf 27}$, corresponding to $H^1(X,V^*)$, are absent. Hence, we are left with families in ${\bf 27}$ multiplets, associated with the cohomology group $H^1(X, V)$. The only type of Yukawa coupling is, therefore, of the form ${\bf 27}\, {\bf 27}\,{\bf 27}$ and it can be calculated from the map~\eqref{cohyuk}. To do this we require polynomial representatives for the two cohomology groups involved, namely for $H^1(X,V)$ and $H^3(X,\wedge^3 V)$.
 
\subsubsection{Polynomial representatives for families in $H^1(X,V)$}
Looking at the long exact sequence in cohomology associated to the short exact monad sequence \eqref{monad}, we find that 
\bea \label{e6fam} 0 &\to&
H^0(X,V) \to H^0(X, B) \stackrel{f}{\longrightarrow} H^0(X, C) \\
\nonumber &\to& H^1(X,V) \to H^1(X, B) \to \ldots \;.  
\eea
For stable  $SU(n)$ bundles we know that $H^0(X,V)=0$. In addition,
if we assume that $H^1(X,B) = 0$, a condition which is always satisfied
for positive monads as a consequence of Kodaira vanishing, it follows that
\bea 
\label{e6fam2} 
H^1(X,V) \cong \frac{H^0(X,C)}{f\left(H^0(X,B)\right)} \ .  
\eea
From Eq.~\eqref{HA}, both cohomology groups on the RHS can be represented in terms of the coordinate ring $A$ of $X$, so we finally have
\begin{equation}
 H^1(X,V) \cong \frac{\bigoplus_{a=1}^{r_C}A_{{\bf c}_a}}{f\left(\bigoplus_{i=1}^{r_B}A_{{\bf b}_i}\right)}\; . \label{H1SU3}
\end{equation} 
The map $f$ in this quotient is induced from the monad map in \eqref{monad}. If we represent the monad map by a matrix $f_{ai}$ of polynomials with multi-degrees ${\bf c}_a-{\bf b}_i$ then its action on a vector of polynomials $(q_i)\in \bigoplus_{i=1}^{r_B}A_{{\bf b}_i}$ is given by
\begin{equation}
 f((q_i))=\left(\sum_{i=1}^{r_B}f_{ai}q_i\right)\in \bigoplus_{a=1}^{r_C}A_{{\bf c}_a}\; . \label{fSU3}
\end{equation} 
This is the action of a  polynomial matrix and it allows us to explicitly compute the polynomial quotient \eqref{H1SU3} once the monad map $f\sim (f_{ai})$ is specified. We note that the degrees of the various polynomials involved are given by the integer vectors ${\bf b}_i$ and ${\bf c}_a$ that define the monad bundle~\eqref{monad}.

We have obtained explicit polynomial representatives for the families and now turn to the ``Yukawa cohomology group'' $H^3(X,\wedge^3 V)$.

\subsubsection{Polynomial representatives for  $H^3(X,\wedge^3 V)$}
Taking the exterior power sequence associated to our monad, as
described in appendix B of reference \cite{Anderson:2008uw}, and splitting it
into short exact sequences we obtain 
\bea 
\nn 0 &\to& \wedge^3 V \to \wedge^3 B \to K_1 \to 0 \\
0 &\to& K_1 \to \wedge^2 B \otimes C \to K_2 \to 0 \\
\nn 0 &\to& K_2 \to B \otimes S^2 C \to S^3 C \to 0\;.  
\eea 
Here we have introduced the (co)-kernels $K_1$ and $K_2$.

The following pieces may be extracted from the associated long-exact
sequences in cohomology.  
\bea 
\label{s1} \ldots &\to& H^2(X, \wedge^3 B) \to H^2(X, K_1) \to 
  H^3(X, \wedge^3 V) \to H^3(X, \wedge^3 B) \to 0 \\ 
\label{s2} \ldots &\to& H^1(X, \wedge^2 B \otimes C) \to H^1(X,
K_2) \to H^2(X, K_1) \to H^2(X, \wedge^2 B\otimes C) \to \ldots
\\ \label{s3} \ldots &\to& H^0(X, B \otimes S^2 C) \to H^0(X, S^3 C)
\to H^1(X, K_2) \to H^1(X, B\otimes S^2 C) \to \ldots 
\eea 
We now assume that the following vanishing conditions 
\bea \label{cond1} 
\nn && H^3(X, \wedge^3 B)=0 \;,\; H^2(X, \wedge^3 B)=0 \\ 
&& H^1(X, \wedge^2 B \otimes C) = 0 \;,\; H^2(X, \wedge^2 B \otimes C)=0 \\ 
\nn && H^1(X, B \otimes S^2 C)=0 \ ,
\eea 
are satisfied. This is automatically the case for positive monad bundles as a consequence of the Kodaira vanishing theorem. Then one can
combine the sequences \eqref{s1}, \eqref{s2} and \eqref{s3} to obtain, 
\bea \label{comparisonclass} \ldots \to
H^0(X, B \otimes S^2 C) \stackrel{F}{\longrightarrow} H^0(X, S^3 C)
\to H^3(X,\wedge^3 V) \to 0 \ .
\eea 
We therefore conclude that
\bea
 \label{missinglabel} 
H^3(X, \wedge^3 V) \cong \frac{H^0(X,S^3C)}{F\left(H^0(X,B\otimes S^2C)\right)}\; .
\eea
Expressing this in terms of the coordinate ring via Eq.~\eqref{HA} as before, leads to
\begin{equation}
 H^3(X, \wedge^3 V) \cong\frac{\bigoplus_{a\geq b\geq c}A_{{\bf c}_a+{\bf c}_b+{\bf c}_c}}{F\left(\bigoplus_{i,a\geq b}A_{{\bf b}_i+{\bf c}_a+{\bf c}_b}\right)}\; . \label{H3SU3}
\end{equation} 
The map $F$ is induced by the monad map $f\sim (f_{ai})$ and, acting on a tensor of polynomials $(q_{iab})\in \bigoplus_{i,a\geq b}A_{{\bf b}_i+{\bf c}_a+{\bf c}_b}$, it can be written as
\begin{equation}
 F((q_{iab}))=\left(\sum_{i=1}^{r_B}q_{i(ab}f_{c)i}\right)\in \bigoplus_{a\geq b\geq c}A_{{\bf c}_a+{\bf c}_b+{\bf c}_c}\; ,
\end{equation}
where the brackets around the indices denote symmetrization. Since $h^3(X,\wedge^3 V)=h^3(X,\cO_X)=1$ we know that this polynomial quotient must be one-dimensional, although this is by no means obvious from the RHS of Eq.~\eqref{H3SU3}. For the  example below we will explicitly verify that this is indeed the case.

\subsubsection{Computing Yukawa couplings}\label{s:method}
From Eq.~\eqref{H1SU3} we know that families are represented by a vector of polynomials $(P_a)_{a=1,\ldots ,r_C}$ with multi-degrees ${\bf c}_a$, subject, of course, to the identifications implied by having to work in the coordinate ring of $X$ and the quotient in Eq.~\eqref{H1SU3}. Let us pick a basis $(P^I_{a})$, in family space, where $I,J,K,\ldots =1,\dots ,h^1(X,V)$ are the family indices. We can then form all possible symmetrized products, $P^I_{(a}P^J_{b}P^K_{c)}$, of these polynomials which are of degree ${\bf c}_a+{\bf c}_b+{\bf c}_c$.  For each choice, $(I,J,K)$, of three families, these products form a three-index symmetric tensor $(P^I_{(a}P^J_{b}P^K_{c)})$ which defines an element of $\bigoplus_{a\geq b\geq c}A_{{\bf c}_a+{\bf c}_b+{\bf c}_c}$ and, hence, from Eq.~\eqref{H3SU3}, an element of the Yukawa cohomology group $H^3(X,\wedge^3 V)$. That the polynomial degrees match in this way is non-trivial and, of course, necessary for our method to work. We can now pick a representative, $(Q_{abc})$, consisting of polynomials with multi-degree ${\bf c}_a+{\bf c}_b+{\bf c}_c$, whose class $[(Q_{abc})]$ spans the quotient~\eqref{H3SU3}. Since we are dealing with a one-dimensional quotient, the class, $[(P^I_{(a}P^J_{b}P^K_{c)})]$, defined by the product of three families, must  be proportional to $[(Q_{abc})]$, so that we can write
\begin{equation}
 \left[(P^I_{(a}P^J_{b}P^K_{c)})\right]=\lambda_{IJK}\left[(Q_{abc})\right]\; . \label{yukdef}
\end{equation} 
The complex numbers $\lambda_{IJK}$ are of course the desired Yukawa couplings. Since the ``comparison class'' $[(Q_{abc})]$ was chosen arbitrarily this only defines the Yukawa couplings up to an overall normalization and, of course, relative to the chosen basis in family space, as expected.

\subsubsection{A simple $E_6$ example} \label{inquisitivebadger}
Let us illustrate this procedure by a simple example on the quintic in $\mathbb{P}^4$. The coordinate ring of the quintic is given by
\begin{equation}
 A=\frac{\mathbb{C}[x_0,\ldots ,x_4]}{\langle p\rangle}\; , \label{Aquintic}
\end{equation}
where $(x_0,\ldots ,x_4)$ are projective coordinates on $\mathbb{P}^4$ and $p$ is the defining quintic polynomial. 
We would like to consider the $SU(3)$ monad bundle defined by
\bea 
\label{eg1} 
0 \to V \to {\cal O}_X(1)^{\oplus 4} \stackrel{f}{\longrightarrow} {\cal O}_X(4) 
\to 0 
\eea 
which is perhaps the simplest positive monad on the quintic. Note that, in this case, the monad map $f$ can be represented by a vector $f=(f_1,\ldots ,f_4)$ of four cubics in $A$. To make contact with the previous general notation, this means that the vectors ${\bf b}_i$ and ${\bf c}_a$ are, in fact, one-dimensional and explicitly given by ${\bf b}_1={\bf b}_3={\bf b}_3 ={\bf b}_4=(1)$ and ${\bf c}_1=(4)$. 

From Eq.~\eqref{H1SU3} it follows that the families are represented by the quotient
\begin{equation}
 H^1(X,V)\cong \frac{A_4}{f\left(A_1^{\oplus 4}\right)} \label{H1quintic}
\end{equation}
of quartic polynomials by the image of four linear polynomials. On a vector $(q_1,\ldots ,q_4)\in A_1^{\oplus 4}$ consisting of four linear polynomials, the map $f$ acts as
\begin{equation}
 f((q_1,\ldots ,q_4))=\sum_{i=1}^4f_iq_i\; .
\end{equation} 
It is easy to count the dimension of this quotient. In general, the
number of degree $k$ polynomials in $n+1$ variables is, \bea
\textnormal{dim} (\mathbb{C}[x_0, \dots, x_n]_k) =
\left(\begin{array}{c}n+k\\n\end{array}\right)\; .  \eea Hence, ${\rm
  dim}A_4=70$ and ${\rm dim} A_1^{\oplus 4}=20$. (In general, one has
to correct for the fact that one is working with the coordinate ring,
rather than the ring of all polynomials. In the present case we are
dividing by an ideal generated by a quintic polynomial so that degrees
$A_k$, where $k<5$ are not affected.) For sufficiently generic choices
of polynomials $f_i$, the map $f$ is injective and we conclude that
the quotient~\eqref{H1quintic} has dimension $70-20=50$. So we are
dealing with a model with $50$ families.

For the Yukawa cohomology group~\eqref{H3SU3} we have in the present case
\begin{equation}
 H^3(X,\wedge^3 V)\cong \frac{A_{12}}{F\left(A_9^{\oplus 4}\right)}\; , \label{H3quintic}
\end{equation} 
where $F$ acts on a vector $(r_1,\ldots ,r_4)\in A_9^{\oplus 4}$ as
\begin{equation}
 F((r_1,\ldots ,r_4))=\sum_{i=1}^4f_ir_i\; .
\end{equation}
As the degrees involved exceed $5$, counting polynomials to determine
the dimension of this quotient is not so simple any more. However, it
is relatively straightforward to extract this information from the
relevant Hilbert series which can be computed with computer algebra
packages such as Macaulay and Singular~\cite{mac,sing}. It turns out
that this dimension is indeed $1$, as it must be from our general
arguments. It should be noted that the computer algebra package
Singular \cite{sing} is fast enough on a standard desktop machine to
perform the calculation of the Yukawa couplings between all 50
families in a matter of minutes. A useful interface for Singular,
designed for use by physicists, may be found here
\cite{StringvacuaBlock}. A sample of the result, for a given choice of
family representatives and monad map, is given in Table~\ref{t:yuk}.

\begin{table}[t]
\begin{center}
\begin{tabular}{ccc}
\begin{tabular}{|c|c|c|c|} \hline
     &    $I=1$&$I=2$&$I=3$\\ \hline
$Y_{11I}$ &0& 0& 0 \\
$Y_{12I}$ &0& 0& $1$\\ 
$Y_{13I}$ &0& $1$& $0$\\ \hline
\end{tabular}
\begin{tabular}{|c|c|c|c|} \hline
     &    $i=1$&$i=2$&$i=3$\\ \hline
$Y_{21I}$ &0& 0& $1$\\
$Y_{22I}$ &0& $0$& $0$\\ 
$Y_{23I}$ &$1$&0& 0\\ \hline
\end{tabular}
\begin{tabular}{|c|c|c|c|} \hline
  &    $i=1$&$i=2$&$i=3$\\ \hline
  $Y_{31I}$ &0& 1& $0$\\ 
  $Y_{32I}$ &$1$& $0$&0\\ 
  $Y_{33I}$ &$0$& 0& 0\\ \hline
\end{tabular}
\end{tabular}
\end{center}
\mycaption{The array of the ${\bf 27}^3$ Yukawa couplings for the $E_6$ GUT associated to the $SU(3)$ monad given in \eqref{eg1} on the quintic. There are 50 families of {\bf 27} multiplets, represented by $H^1(X,V)$; we select three of these for illustrative purposes here, as indexed by $I=1,2,3$. These are represented by the monomials $x_4^4$, $x_2^2 x_3^2$ and $x_0^2 x_1^2$ respectively (that is, these are the normal forms of the equivalence class of polynomials representing these families). The normal form of the comparison class in this calculation was $x_0^2 x_1^2 x_2^2 x_3^2 x_4^4$. The monad map is given by $f=(x_0^3,x_1^3,x_2^3,x_3^3)$.}\label{t:yuk}
\end{table}

\subsection{$SU(4)$ vector bundles and $SO(10)$ GUTs} \label{mrso10}
Having introduced our general method of computing Yukawa couplings for
the case of $SU(3)$ bundles, let us move on to consider the case of
$SU(4)$ bundles. From Table~\ref{t:break} we have the following symmetry
breaking pattern and decomposition of the matter field representations
\bea
E_8 &\supset& SU(4) \times SO(10) \\
{\bf 248} &=& ({\bf 15},{\bf 1}) \oplus ({\bf 1},{\bf 45}) \oplus
({\bf 4},{\bf 16}) \oplus (\overline{\bf 4},\overline{\bf 16}) \oplus
({\bf 6},{\bf 10})\; .  \eea The $({\bf 15},{\bf 1})$ term corresponds
to bundle moduli that are counted by the cohomology group $H^1(X, V
\otimes V^*)$. As in the $E_6$ case anti-families in $(\overline{\bf
  4},\overline{\bf 16})$ multiplets are absent for positive
monads. Therefore, the relevant Yukawa couplings are of the form ${\bf
  10}\,{\bf 16}\, {\bf 16}$ and couple two families in ${\bf 16}$
multiplets, associated to the cohomology group $H^1(X,V)$, to a Higgs
multiplet in ${\bf 10}$, associated to the cohomology group $H^1(X,
\wedge^2 V)$. The associated Yukawa coupling can be computed by
considering the map \eqref{so10cohyuk}, so we need polynomial
representations for $H^1(X, V)$ and $H^2(X, \wedge^2 V)$.

\vspace{0.1cm}

Polynomial representatives for the families in $H^1(X,V)$ can be worked out in exactly the same way as for the $E_6$ case and the result is given by Eqs.~\eqref{H1SU3} and \eqref{fSU3}.

\subsubsection{Polynomial representatives for Higgs multiplets in $H^1(X, \wedge^2V)$}\label{so10higgs}
For the ${\bf 10}$ multiplets, corresponding to $H^1(X, \wedge^2 V)\simeq H^2(X,\wedge^2 V)$, we introduce an exterior
power sequence associated to the defining sequence of the monad, \eref{monad}. Splitting the sequence up using the (co)-kernel $K_3$ we obtain the
following
\bea
\nonumber 0 &\to& \wedge^2 V \to \wedge^2 B \to K_3 \to 0 \\
0 &\to& K_3 \to B \otimes C \to S^2 C \to 0 \ .
\eea
These induce the following long exact sequences in cohomology
\bea 
\dots &\to& H^1(X,\wedge^2 B) \to H^1(X, K_3) \to H^2(X, \wedge^2 V) \to H^2(X, \wedge^2
B) \to \ldots\\
\label{secondso10}
\ldots &\to& H^0(X, B \otimes C) \stackrel{F}{\longrightarrow} H^0(X, S^2 C)
\to H^1(X, K_3) \to H^1(X, B \otimes C) \to \ldots \; . 
\eea 
Thus, if $H^1(X, \wedge^2 B) = H^2(X, \wedge^2 B)=0$, which are two of our vanishing conditions in Table~\ref{conditions} satisfied for all positive monads, we have that $H^1(X, K_3) \cong H^2(X, \wedge^2V)$. Together with the vanishing condition $H^1(X, B \otimes C) \cong 0$, again satisfied for all positive monads, this can be used in \eqref{secondso10} to obtain
\bea
 \label{thirdso10}
  H^1(X, \wedge^2 V) \cong\frac{H^0(X,S^2 C)}{F\left(H^0(X,B\otimes C)\right)}\; .
\eea
From Eq.~\eqref{HA} this translates to
\begin{equation}
 H^1(X, \wedge^2 V) \cong\frac{\bigoplus_{a\geq b}A_{{\bf c}_a+{\bf c}_b}}{F\left(\bigoplus_{i,a}A_{{\bf b}_i+{\bf c}_a}\right)}\; .
 \label{HSU4}
\end{equation} 
The map $F$ is induced by the monad map $f$ and, acting on a tensor of polynomials $(q_{ia})\in \bigoplus_{i,a}A_{{\bf b}_i+{\bf c}_a}$, it can be written as
\begin{equation}
 F((q_{ia}))=\left(\sum_{i=1}^{r_C}q_{i(a}f_{b)i}\right)\in \bigoplus_{a\geq b}A_{{\bf c}_a+{\bf c}_b}\; . \label{Hmap}
\end{equation}

\subsubsection{Computing Yukawa couplings}
We would now like to compute Yukawa couplings by mapping in the way indicated in \eqref{so10cohyuk}. We note that, from Eq.~\eqref{H1SU3}, a basis in family space takes the form $(P^I_{a})$, where $I,J,K,\ldots = 1,\dots ,h^1(X,V)$ are family indices, and the polynomials are of multi-degree ${\bf c}_a$. A basis for the Higgs space~\eqref{HSU4} can be expressed in terms of multi-degree ${\bf c}_a+{\bf c}_b$ polynomials $(H^A_{ab})$, where $A=1,\dots ,h^1(X,\wedge^2 V)$ numbers the Higgs multiplets and $(ab)$ is a symmetrized index pair. Hence, the product of two polynomials representing families is precisely of the right multi-degree to be interpreted as an element of the Higgs polynomial space. We can, therefore, write
\begin{equation}
 \left[(P^I_{(a}P^J_{b)})\right]=\sum_A\lambda_{AIJ}\left[(H^A_{ab})\right] \label{SO10yuk}
\end{equation}
with $\lambda_{AIJ}$ being the desired Yukawa couplings. An explicit example with just one Higgs multiplet will be discussed in the next section.

\subsection{$SU(5)$ vector bundles and $SU(5)$ GUTs}
The final case we shall consider is that of $SU(5)$ bundles. For this
case we have the following symmetry breaking pattern and decomposition
of the matter field representations (to avoid confusion we have marked
the GUT $SU(5)$ group with a subscript ${\rm GUT}$): \bea
E_8 &\supset& SU(5) \times SU(5)_{\rm GUT} \\
{\bf 248} &=& ({\bf 24},{\bf 1}) \oplus ({\bf 1},{\bf 24}) \oplus
({\bf 5},{\bf 10}) \oplus (\overline{\bf 5},\overline{\bf 10}) \oplus
( \overline{\bf 10},{\bf 5}) \oplus ( {\bf 10}, \overline{\bf 5}) \;.\eea
The absence of anti-generations for positive monad bundles implies
that the $(\overline{\bf 5},\overline{\bf 10})$ states in the
decomposition above are not present in the low energy spectrum as
$H^1(X,V^*)=0$. The relevant Yukawa couplings are then of the two
types ${\bf 5}\, {\bf 10}\,{\bf 10}$ and ${\bf 10}\,\overline{\bf
  5}\,\overline{\bf 5}$ and they have to be computed from the
maps~\eqref{cohyuk5} and \eqref{cohyuklast}. This means we must have
polynomial representations for $H^1(X, V)$, $H^1(X, \wedge^2 V)$,
$H^2(X, \wedge^4 V)$ and $H^2(X, \wedge^2V)$.

\vspace{0.1cm}

We start, as in the other cases, by obtaining representatives for the cohomologies associated to the
families residing in ${\bf 10}$ multiplets. They correspond to the cohomology group $H^1(X,V)$ and
can be dealt with in exactly the same way as the ${\bf 16}$ multiplets in the $SO(10)$ case and
the ${\bf 27}$ multiplets for $E_6$. Hence, their polynomial representatives are given by Eqs.~\eqref{H1SU3} and \eqref{fSU3}.

\subsubsection{Polynomial representatives for $H^1(X, \wedge^2 V)$}
Polynomial representatives for the $\overline{\bf 5}$ multiplets in $H^1(X, \wedge^2 V)$ may be obtained as for the ${\bf 10}$ multiplets in the $SO(10)$ case, see Section~\ref{so10higgs}. However, in the present case these particular representatives are not suitable for a calculation of Yukawa couplings following Eq.~\eqref{cohyuklast} since they do not square to the polynomial representatives for $H^2(X,\wedge^4 V)$, as determined below. We, therefore, have to follow a slightly more complicated approach. As usual, we use an exterior power sequence associated to the defining sequence of the monad. Splitting this sequence up, using the (co)-kernel $K_4$, we obtain the two short exact sequences
\bea \label{so10comp1}
&& 0 \to
\wedge^2 V \to \wedge^2 B \to K_4 \to 0 \\ \nonumber && 0 \to K_4 \to
B \otimes C \to S^2 C \to 0 \ .
\eea 
The corresponding long exact sequences in cohomology contain the parts
\bea 
\nonumber && \ldots \to H^0(X,
\wedge^2 B) \stackrel{f_1}{\longrightarrow} H^0(X, K_4) \to H^1(X,
\wedge^2 V) \to H^1(X, \wedge^2 B) \to \ldots \\ && 0 \to H^0(X, K_4)
\to H^0(X, B \otimes C) \stackrel{f_2}{\longrightarrow} H^0(X, S^2 C)
\to \ldots \; .
\eea 
Given that $H^0(X, K_4) \cong \textnormal{Ker}(f_2)$ it follows that $f_2\circ f_1=0$ and with the polynomial representatives
\begin{eqnarray}
 H^0(X,\wedge^2 B)&\cong&\bigoplus_{i>j}A_{{\bf b}_i+{\bf b}_j}\\
 H^0(X,B\otimes C)&\cong&\bigoplus_{i,a}A_{{\bf b}_i+{\bf c}_a}\\
 H^2(X,S^2 C)&\cong&\bigoplus_{a\geq b}A_{{\bf c}_a+{\bf c}_b}
\end{eqnarray} 
we have the complex
\begin{equation}
 \bigoplus_{i>j}A_{{\bf b}_i+{\bf b}_j}\stackrel{f_1}{\longrightarrow}\bigoplus_{i,a}A_{{\bf b}_i+{\bf c}_a}\stackrel{f_2}{\longrightarrow}\bigoplus_{a\geq b}A_{{\bf c}_a+{\bf c}_b}\; .
\end{equation} 
On polynomial tensors $(Q_{ij})$ and $(q_{ia})$ the two maps above act as
\begin{equation}
 f_1((Q_{ij}))=\left(\sum_jf_{ai}Q_{ij}\right)\; ,\quad f_2((q_{ia}))=\left(\sum_iq_{i(a}f_{b)i}\right)\; ,
\end{equation}
which confirms explicitly that $f_2\circ f_1=0$. The desired bundle
cohomology $H^1(X,\wedge^2 V)$ is now given, if $H^1(X, \wedge^2
B)=0$, by the cohomology of the above complex, that is,
\begin{equation}
 H^1(X,\wedge^2 V)\simeq \frac{{\rm Ker}(f_2)}{{\rm Im}(f_1)}\; . \label{su5higgsb}
\end{equation}

\subsubsection{Polynomial representatives for $H^2(X, \wedge^4 V)$}
Let us now obtain an appropriate polynomial description for the ${\bf 10}$ multiplets in  $H^2(X, \wedge^4 V)$ as required for calculating the ${\bf 10}\,\overline{\bf 5}\,\overline{\bf 5}$ Yukawa couplings from Eq.~\eqref{cohyuklast}. Consider
the exterior power sequence of the monad exact sequence, split by introducing (co)-kernels $K_5$,$K_6$ and $K_{7}$.
\bea\label{wedge4} && 0 \to \wedge^4 V \to \wedge^4 B \to K_5 \to 0 \\
&& 0 \to K_5 \to \wedge^3 B \otimes C \to K_6 \to 0\label{wedge4b} \\ 
&& 0 \to K_6 \to \wedge^2 B \otimes S^2 C \to K_{7} \to 0\label{wedge4c} \\ 
&& 0 \to K_{7} \to B \otimes S^3 C \to S^4 C \to 0\label{wedge4d} \eea 
For our argument we require the following parts of the associated long exact sequences.
\begin{eqnarray}
  \ldots& \to& H^1(X,\wedge^4B)\to H^1(X, K_5) \to H^2(X,\wedge^4 V) \to  H^2(X, \wedge^4 B) \to \ldots\\
  \ldots &\to& H^0(X, \wedge^3 B \otimes C) \stackrel{f_3}{\longrightarrow} H^0(X, K_6) \to H^1(X, K_5) \to H^1(X, \wedge^3 B \otimes C) \to \ldots \label{les2}\\
  0 &\to& H^0(X, K_6)\to H^0(X, \wedge^2 B \otimes S^2 C) \stackrel{f_4}{\longrightarrow}  H^0(X, K_{7}) \to   \ldots\\
  0 &\to& H^0(X, K_{7}) \to H^0(X, B \otimes S^3 C) \to  H^0(X,S^4 C) \to \ldots\label{les4}
\end{eqnarray}
From our vanishing assumptions, which we remind the reader are
automatically satisfied by the positive monads, $H^1(X,\wedge^4
B)=H^2(X,\wedge^4B)=0$ and, hence, the first of these sequences
implies that $H^2(X,\wedge^4 V)\cong H^1(X,K_5)$. The last two
sequences tell us that $H^0(X,K_6)$ injects into
$H^0(X,\wedge^2B\otimes S^2C)$ and $H^0(X,K_{7})$ injects into
$H^0(X,B\otimes S^3C)$. Introducing the polynomial representatives
\begin{eqnarray}
 H^0(X,\wedge^3B\otimes C)&\cong& \bigoplus_{i>j>k,a}A_{{\bf b}_i+{\bf b}_j+{\bf b}_k+{\bf c}_a}\\
 H^0(X,\wedge^2B\otimes S^2C)&\cong&\bigoplus_{i>j,a\geq b}A_{{\bf b}_i+{\bf b}_j+{\bf c}_a+{\bf c}_b}\\
 H^0(X,B\otimes S^3C)&\cong&\bigoplus_{i,a\geq b\geq c}A_{{\bf b}_i+{\bf c}_a+{\bf c}_b+{\bf c}_c}\; ,
\end{eqnarray} 
we can therefore combine \eqref{les2}--\eqref{les4} to form the complex
\begin{equation}
 \bigoplus_{i>j>k,a}A_{{\bf b}_i+{\bf b}_j+{\bf b}_k+{\bf c}_a}\stackrel{f_3}{\longrightarrow}\bigoplus_{i>j,a\geq b}A_{{\bf b}_i+{\bf b}_j+{\bf c}_a+{\bf c}_b}\stackrel{f_4}{\longrightarrow}\bigoplus_{i,a\geq b\geq c}A_{{\bf b}_i+{\bf c}_a+{\bf c}_b+{\bf c}_c}\; .
\end{equation} 
On polynomial tensors $(Q_{ijka})$ and $(q_{ijab})$ the above maps $f_3$ and $f_4$ acts as
\begin{equation}
 f_3((Q_{ijka}))=\sum_kQ_{ijk(a}f_{b)k}\; ,\quad f_4((q_{ijab}))=\sum_jq_{ij(ab}f_{c)j}\; .
\end{equation}
As before, the desired cohomology $H^2(X,\wedge^4 V)$ is given, if
$H^1(X, \wedge^3 B \otimes C)=0$, by the cohomology of this complex,
that is,
\begin{equation}
  H^2(X,\wedge^4 V)\cong\frac{{\rm Ker}(f_4)}{{\rm Im}(f_3)}\; . \label{su5alt}
\end{equation}  

\subsubsection{Polynomial representatives for $H^2(X, \wedge^2 V)$}
Finally, we require polynomials to represent the ${\bf 5}$ multiplets in $H^2(X, \wedge^2 V)$, to calculate the ${\bf 5}\, {\bf 10}\,{\bf 10}$ Yukawa couplings from Eq.~\eqref{cohyuk5}. We once again consider the long exact sequence in cohomology
induced by \eqref{so10comp1}. This contains the following pieces:
\bea 
\ldots &\to& H^1(X, \wedge^2 B) \to H^1(X, K_4) \to H^2(X, \wedge^2
V) \to H^2(X,\wedge^2 B) \to \ldots \\
\ldots &\to& H^0(X, B \otimes C) \stackrel{f_5}{\longrightarrow} H^0(X,
S^2 C) \to H^1(X, K_4) \to H^1(X, B \otimes C) \to \ldots 
\eea 
Given the vanishing assumptions $H^1(X, \wedge^2 B) = H^2(X, \wedge^2 B) = H^1(X, B \otimes C) =0$, which are automatically satisfied for positive monads, the first of these sequences implies that $H^2(X,\wedge^2 V)\cong H^1(X,K_4)$.
Using this in the second sequence leads to
\begin{equation}
 H^2(X,\wedge^2 V)\cong\frac{H^0(X,S^2C)}{f_5\left(H^0(X,B\otimes C)\right)}\; .
\end{equation}
Written in terms of polynomial representatives this means
\begin{equation}
  H^2(X,\wedge^2 V)\cong\frac{\bigoplus_{a\geq b}A_{{\bf c}_a+{\bf c}_b}}{f_5\left(\bigoplus_{i,a}A_{{\bf b}_i+{\bf c}_a}\right)}\; .
  \label{su5higgs}
\end{equation}  
On polynomial tensors $(q_{ia})$ the map $f_5$ acts as
\begin{equation}
 f_5((q_{ia}))=\left(\sum_iq_{i(a}f_{b)i}\right)\; .
\end{equation}  

\subsubsection{Computing Yukawa couplings}
We begin by summarising the polynomial representations for the various multiplets. For the families in ${\bf 10}$ multiplets we have a basis of polynomials $(P_a^I)$ with multi-degrees ${\bf c}_a$, where $I,J\dots =1,\ldots ,h^1(X,V)$, as before.  From Eq.~\eqref{su5higgs}, ${\bf 5}$ multiplets are represented by multi-degree ${\bf c}_a+{\bf c}_b$ polynomials $(H_{ab}^A)$, where $A,B,\dots = 1,\ldots ,h^1(X,\wedge^3 V^\star)$ and $(ab)$ is a symmetric index pair. Eq.~\eqref{su5higgsb} shows that $\overline{\bf 5}$ multiplets can be represented by polynomials $(\bar{H}^{\bar A}_{ia})$ of multi-degree ${\bf b}_i+{\bf c}_a$, where $\bar{A},\bar{B},\dots =1,\ldots ,h^1(X,\wedge^2 V)$. Finally, from Eq.~\eqref{su5alt} we have an alternative polynomial representation for the families in ${\bf 10}$ by multi-degree ${\bf b}_i+{\bf b}_j+{\bf c}_a+{\bf c}_b$ polynomials $(\tilde{P}^I_{ijab})$, where $(ij)$ is an anti-symmetric and $(ab)$ a symmetric index pair.

Given these polynomial representatives, the ${\bf 5}\, {\bf 10}\, {\bf 10}$ Yukawa couplings $\lambda_{AIJ}$ and the ${\bf 10}\,\overline{\bf 5}\,\overline{\bf 5}$ Yukawa couplings $\lambda_{I\bar{A}\bar{B}}$ can be computed from
\begin{eqnarray}
 \left[(P_{(a}^IP_{b)}^J)\right]&=&\sum_A\lambda_{AIJ}\left[(H_{ab}^A)\right]\\
\left[(\bar{H}^{\bar A}_{[i|(a}\bar{H}^{\bar B}_{|j]b)})\right]&=&\sum_I\lambda_{I\bar{A}\bar{B}}\left[(\tilde{P}_{ijab}^I)\right]\; .
\end{eqnarray} 
This concludes our general discussion. We now move on to give a comprehensively worked example of some physical
interest in the $SO(10)$ case.

%
\section{An example: One Higgs  multiplet and one heavy family}
\label{bigeg}
\setall
\subsection{The model}\label{bigegmodel}
As in our previous example, in section \ref{inquisitivebadger}, we
consider the quintic in $\mathbb{P}^4$. The coordinate ring is given
by
\begin{equation}
 A=\frac{\mathbb{C}[x_0,\ldots ,x_4]}{\langle p\rangle}\; ,
\end{equation}
where $(x_0,\ldots ,x_4)$ are the projective coordinates on $\mathbb{P}^4$ and $p$ is the defining quintic polynomial. 
In this section, we will consider the following monad on the quintic.
\bea 
\label{examplemonad} 0 \to V \to {\cal O}_X(1)^{\oplus 7}
\stackrel{f}{\longrightarrow} {\cal O}_X(2)^{\oplus 2} \oplus {\cal
  O}_X(3) \to 0 \ . 
\eea 
This short exact sequence defines an $SU(4)$ bundle and thus we are discussing an $SO(10)$ GUT theory as in \S\ref{mrso10}. The Yukawa couplings we shall calculate for this model are thus of the form ${\bf 10}\,{\bf 16}\,{\bf 16}$.
The monad map $f$ can be written as $f=(f_{1i},f_{2i},f_{3i})$, where $i=1,\ldots ,7$ runs over the seven
$\cO_X(1)$ line bundles and $f_{1i}$, $f_{2i}$ are degree one polynomials in $A$ while $f_{3i}$ are degree two polynomials. From the general discussion in \S\ref{mrso10} it follows that the families in $H^1(X,V)$ can be represented by polynomials as
\begin{equation}
 H^1(X,V)\cong \frac{A_2^{\oplus 2}\oplus A_3}{f\left(A_1^{\oplus 7}\right)}\; . \label{famex}
\end{equation}
Given that ${\rm dim} A_2=15$, ${\rm dim} A_3=35$ and ${\rm dim} A_1=5$, and choosing the map $f$ sufficiently general so it is injective, it follows that this model has $30$ families.  From Eq.~\eqref{HSU4}, the Higgs multiplets in $H^1(X,\wedge^2 V)$ can be represented by
\begin{equation}
 H^1(X,\wedge^2 V)\cong\frac{A_4^{\oplus 3}\oplus A_5^{\oplus 2}\oplus A_6}{F\left(A_3^{\oplus 14}\oplus A_4^{\oplus 7}\right)}\; , \label{Hex}
\end{equation} 
where the map $F$ has been defined in Eq.~\eqref{Hmap}. If we write polynomials in the denominator of this quotient as $(q_{(3)i},\tilde{q}_{(3)i},q_{(4)i})^T$, where $i=1,\ldots ,7$ and the first index indicates the polynomial degree, and polynomials in the numerator as $(Q_{(4)1},Q_{(4)2},Q_{(5)1},Q_{(4)3},Q_{(6)},Q_{(5)2})^T$, with the first index again indicating the polynomial degree, then this map can be explicitly written as
\bea 
\label{map} 
F\left( \ba{c} q_{(3) i} \\
  \tilde{q}_{(3)i} \\ q_{(4)i} \ea \right)=\left( \ba{c} Q_{(4)1} \\ Q_{(4)2} \\ Q_{(5)1} \\ Q_{(4)3} \\ Q_{(6)} \\
  Q_{(5)2}\ea \right) = \left( \ba{ccc}
  f_{1i} & 0 & 0 \\
  \frac{1}{2} f_{2i} & \frac{1}{2} f_{1i} & 0 \\
  \frac{1}{2} f_{3i} & 0 & \frac{1}{2} f_{1i} \\
  0 & f_{2i} & 0 \\
  0 & 0 & f_{3i} \\
  0 & \frac{1}{2} f_{3i} & \frac{1}{2} f_{2i} \ea \right) \left( \ba{c} q_{(3) i} \\
  \tilde{q}_{(3)i} \\ q_{(4)i} \ea \right) \ .  \eea We note that this
is a $6\times 21$ matrix of polynomials. We can use this explicit map
to compute the dimension of the quotient~\eqref{Hex}. For generic
choices of the monad map $f$ it turns out that this dimension is zero,
so there are no Higgs multiplets. This confirms the general result,
found in references \cite{Anderson:2007nc,Anderson:2008uw}, that
$h^1(X,\wedge^2 V)=0$ at a generic point in bundle moduli space. This
generic case is of course of no interest in our context since Yukawa
couplings of the form ${\bf 10}\,{\bf 16}\, {\bf 16}$ are not present.

\subsection{Engineering Higgs multiplets}
\label{s:1Higgs}
To arrive at physically more interesting cases we have to understand
how to engineer models with one (or possibly more than one) Higgs
multiplet. This is typically not easy from a technical point of view
and it was a particular challenge in the effort to find the exact MSSM
spectrum from heterotic compactifications based on
elliptically-fibered Calabi-Yau
manifolds~\cite{Braun:2005ux,Donagi:2004qk}. In the present framework,
it is at least straightforward to state what needs to be done in
principle. We need to make special choices for the polynomials
defining the monad map $f$ in such a way that the induced map $F$ in
Eq.~\eqref{map} leads to dimension-one quotient~\eqref{Hex}. At the
same time, $f$ still has to be sufficiently general so that $V$, as
defined by the monad short exact sequence, is indeed a bundle rather
than merely a sheaf.

To examine this in detail we can consider $F$ as a map between modules
$F: A(-3)^{\oplus 14}\oplus A(-4)^{\oplus 7}\longrightarrow
A(-4)^{\oplus 3}\oplus A(-5)^{\oplus 2}\oplus A(-6)$ and then examine
the Hilbert function of ${\rm Coker} (F)$ at degree zero. As stated
above, for generic choices of $f$, that is, at generic points in
bundle moduli space, the Hilbert function at degree zero
vanishes. Another way of stating the same fact is that the Hilbert
functions of the ideals \bea \label{ideals} \langle f_{1i}\rangle\;,\;
\langle f_{2i}, f_{1i}\rangle \;,\;\langle f_{3i},f_{1i} \rangle\;,\;
\langle f_{2i} \rangle\;,\; \langle f_{3i}\rangle \;,\; \langle
f_{3i},f_{2i}\rangle \;.  \eea of the Calabi-Yau's coordinate ring,
that correspond to the images of the matrix rows in~\eqref{map}, are
each individually zero at the appropriate degrees, that is at degrees
$(4,4,5,4,6)$. This suggests a simple way of engineering one Higgs
multiplet. Rather than dealing with the full complication of the
map~\eqref{map} and its associated Hilbert function, we can focus on
one row and produce a dimension one entry at the appropriate degree of
the associated ideal, while keeping the dimensions zero for all other
ideals. In particular, we can specialise the polynomials $f_{1i}$ so
that the ideal $\langle f_{1i}\rangle$ has dimension one at degree
$4$. Since all of the other ideals depend on polynomials other than
$f_{1i}$, one finds, upon doing this, that the Hilbert function for the
remaining ideals can be kept zero at the appropriate degrees. As a
result, the dimension of the quotient~\eqref{Hex} is one and we have
engineered an example with one Higgs multiplet.

We still have to check that there exists a choice for $f$ along the
above lines that defines a bundle rather than just a sheaf. To do
this we consider the explicit example \bea
\label{engineered} 
\nonumber 
f_{1i} &=& (40 x_3 + 94 x_4, 117 x_3 + 119 x_4, 449 x_3
+ 464 x_4 + 266 x_0 + 195 x_1 + 173 x_2, \\ 
&& 
\qquad
306 x_2, 273 x_3, 259 x_3 + 291 x_4,76 x_3 + 98 x_2) \;, 
\eea
with the remaining polynomials in $f$ being left generic. This choice has been engineered in the way described above and it can be verified that it indeed leads to precisely one Higgs multiplet. In addition, one can check that the locus in $\mathbb{P}^4$ where the polynomial matrix $f$ degenerates (that is, where its rank is not maximal) does not intersect a sufficiently general quintic and, hence, leads to a bundle on the quintic (although not to a bundle on $\mathbb{P}^4$).

\subsection{The mass matrix}

We now wish to calculate the Yukawa couplings in this class of  examples with one Higgs multiplet. To do so we first pick $30$ family representatives $P^I=(P^I_1,P^I_2,P^I_3)\in A_2^{\oplus 2}\oplus A_3$ whose associated classes form a basis of \eqref{famex}. Further we choose a Higgs representative $H=(H_1,\ldots ,H_6)\in A_4^{\oplus 3}\oplus A_5^{\oplus 2}\oplus A_6$ whose class spans the one-dimensional space~\eqref{Hex}. The Yukawa couplings then follow from Eq.~\eqref{SO10yuk} and form a symmetric matrix $\lambda_{IJ}$. Given that we do not know the matter field kinetic terms, the only physically significant property of this matrix is its rank. It turns out, with the map~\eqref{engineered} this rank is precisely one. 
\begin{quote}
\emph{The monad in \eqref{examplemonad} gives rise to one massive family in 
four dimensions at the point specified by \eqref{engineered} in its bundle moduli space}. 
\end{quote}
In fact, this structure is somewhat more generic. Let us consider, more generally, an $SO(10)$ model with a basis $\{P_A\}$ of  ${\cal F}\equiv\bigoplus_aA_{{\bf c}_a}$, such that $\{P_I\}\subset\{P_A\}$ is a set of family representatives and a single Higgs multiplet represented by $H\in{\cal H}\equiv\bigoplus_{a\geq b}A_{{\bf c}_a+{\bf c}_b}$. Note that ${\cal H}=S^2{\cal F}$, so the symmetric tensor products $\{P_A\otimes_SP_B\}$ form a basis of ${\cal H}$ and we can introduce a hermitian scalar product, $\langle\,\cdot\,,\,\cdot\,\rangle$, on ${\cal H}$ such that this basis is orthonormal. From Eq.~\eqref{SO10yuk} it then follows that the Yukawa matrix is given by the scalar product $\lambda_{IJ}\sim\langle P_I\otimes_SP_J,H\rangle$. The Higgs representative $H$ can, of course, always be written as a linear combination $H=\sum_{A,B}H_{AB}P_A\otimes_SP_B$. Inserting this into the above scalar product expression for the Yukawa couplings one finds that
\begin{equation}
 \lambda_{IJ}\sim H_{IJ}\; .
\end{equation}
This means, in a case where the Higgs representative can be expressed
in terms of the family representatives, so that
$H=\sum_{I,J}H_{IJ}P_I\otimes_SP_J$, the Yukawa matrix and the matrix
representing the Higgs are proportional. In particular, their rank has
to be the same. The method of engineering models with one Higgs
multiplet described above typically leads to a Higgs representative
which can be written as the square of a vector $v_I$, that is
$H=\sum_{I,J}v_Iv_JP_I\otimes_SP_J$ \footnote{We would like to thank
  Tony Pantev for very helpful comments on this point.}. This can be
seen as follows.

Let us choose a Higgs representative by taking a so called ``normal
form'' of a sufficiently generic linear combination of terms,
$\sum_{A,B} c_{A, B} P_A \otimes_S P_B$ where $c_{A,B}$ are some
randomly generated coefficients. We take this normal form by
performing the Buchberger Algorithm \cite{Buchberger,talk} on the
linear combination relative to the module generated by the map
polynomials defining the one dimensional class \eqref{Hex}. Given our
method of engineering a single Higgs, as discussed in the previous
sub-section, the resulting Higgs representative will be of the form
$\left( Q_{(4)1},0,0,0,0,0 \right)^T$. An inspection of the Buchberger
algorithm \cite{Buchberger,talk} reveals that $Q_{(4)1}$ in this
expression will be a single monomial. It is in fact the ``lagging
monomial'' of degree four that is not in the ideal $\left< f_{1i}
\right>$. That is, it is the degree four monomial that is lowest
according to the monomial ordering used in the Buchberger algorithm,
which does not appear as an element of $\left< f_{1i} \right> \subset
A$. If this lagging monomial is a square, then clearly our Higgs
representative is the square of a family representative. This is
always the case for the type of example considered in Sections
\ref{bigegmodel} and \ref{s:1Higgs}. The $f_{1i}$ are all linear
polynomials for the monad given in \eqref{examplemonad}. Given this,
the lagging monomial is some variable to the fourth power and $H = c
\; (x_i^2,0,0)^T \otimes_S (x_i^2,0,0)^T$ for some $i$ and some
constant $c$.

As a result, the matrix associated to the Higgs representative and,
hence, the Yukawa matrix has rank one. We see that there is a close
relation between our method for engineering one-Higgs models and
obtaining precisely one heavy family. In conclusion we can state the
following.
\begin{quote}
\emph{A model in which one
Higgs is engineered in the manner described in \S\ref{s:1Higgs} will
generically have one heavy family.}
\end{quote}

\section{Conclusions}
\setall
In this paper we have introduced a simple algorithm for calculating
Yukawa couplings in a wide class of heterotic models. The
compactifications we have considered are on smooth Calabi-Yau spaces and are
not restricted to the standard embedding. The methods can be used to calculate Yukawa couplings
for a large class of bundles on complete-intersection Calabi-Yau manifolds. Such a systematic procedure
for non-standard embedding models has not been presented in the literature before and, we believe, 
constitutes substantial advance.

The key to our methodology is to obtain polynomial representatives for
family and other relevant multiplets whose degrees are compatible with
one another.  In practice, this requires finding polynomial
representatives for various cohomology groups whose degrees are such
that our procedure of polynomial multiplication and reduction to
normal form may be carried out. Because of the simple, algebraic,
nature of the resulting algorithm, the calculations can be carried out
on a computer and we have done this in the text for a number of
concrete examples.

We should stress again that we have calculated the
\emph{superpotential} contributions to the Yukawa couplings. The
K\"ahler potential for the matter fields remains an unknown quantity
in these non-standard embedding models. Nevertheless we have shown how
some physically relevant information can be extracted from our results
by focusing on quantities, such as the rank in the case of a Yukawa
matrix, which are unaffected by the choice of basis in family space.

The final example, presented in section \ref{bigeg}, demonstrates the
power of these methods. This is an example of a smooth Calabi-Yau
compactification leading to an $SO(10)$ GUT. We have shown
how one may isolate loci in bundle moduli space where the model has
precisely one Higgs multiplet, residing in the ${\bf 10}$ representation.
Our approach based on polynomial representatives makes this conceptually
rather simple and merely requires making specific choices for the polynomials defining 
the bundle. In practice, it is not always straightforward to find these but we have
described a simple method to engineer viable cases. 

We have then shown that the structure of this one-Higgs model leads to
a Yukawa matrix of rank one, and so to precisely one massive family.
Moreover, the relation between our method of engineering one Higgs
multiplet and obtaining one massive family seems to be more general,
an observation which might be important for building heterotic models
with a phenomenologically viable pattern of fermion masses.

While our method of computing Yukawa couplings by multiplying
polynomial representatives has been presented in the context of a
particular class of models, the underlying mathematical structure --
the sheaf-module correspondence -- is quite general and we expect
related methods to work for other Calabi-Yau and bundle
constructions. This work should be of considerable utility in checking
conclusions about the vanishing of Yukawa couplings resulting from the
research presented in references
\cite{Anderson:2009sw,Longerpaper}. Eventually, one would like to
calculate Yukawa couplings in the context of more realistic models,
where the GUT symmetry is broken due to Wilson lines. We expect that
the methods described in this paper can be readily applied to such
models, basically by projecting onto the various equivariant
sub-spaces of the cohomology groups involved.

\section*{Acknowledgements}
We gratefully acknowledge enlightening discussions with Philip
Candelas and Tony Pantev.  D.~G.~is partially supported by NSF grants
DMS 08-10948 and DMS 03-11378, L.~A.~is supported by the DOE under
contract No. DE-AC02-76-ER-03071, J.~G.~, by STFC, UK, Y.-H.~H.~, by
an Advanced Fellowship from the STFC, UK, as well as the FitzJames
Fellowship of Merton College, Oxford, and A.~L.~, by the EC 6th
Framework Programme MRTN-CT-2004-503369. L.~A.~, J.~G.~and A.~L.~would
like to thank the organisers of the 2008 Vienna ESI workshop
``Mathematical Challenges in String Phenomenology'' where part of this
work was completed. D.~G.~would like to thank Y.-H.~H.~for hospitality
at the University of Oxford and Merton College, Oxford.

\section*{\Huge Appendices}
\appendix

\section{Koszul Complex and Polynomial Representatives}
\label{ap:koszul}
\setall
In this appendix we justify the relation \eqref{HA} between sections of line bundles on a CICY manifold and its coordinate ring. First let us recall the general set-up and the notation. We work in an ambient space $\cA=\mathbb{P}^{n_1}\times\dots\times\mathbb{P}^{n_m}$ with projective coordinates $({\bf x}^{(1)},\ldots ,{\bf x}^{(m)})$. Line bundles on $\cA$ are denoted by $\cO_\cA({\bf k})=\cO_{\mathbb{P}^{n_1}}(k_1)\otimes\dots\otimes \cO_{\mathbb{P}^{n_m}}(k_m)$, where ${\bf k}=(k_1,\ldots ,k_m)$. The associated ring 
\begin{equation}
 R=\mathbb{C}[{\bf x}^{(1)},\ldots ,{\bf x}^{(m)}]
\end{equation}
is multi-graded by an $m$-dimensional grade vector ${\bf k}=(k_1,\ldots ,k_m)$ where $k_r$ specifies the degree in the projective coordinates ${\bf x}^{(r)}$ of $\mathbb{P}^{n_r}$.  The multi-degree ${\bf k}$ part of $R$ is denoted by $R_{\bf k}$. Sections $H^0(X,\cO_\cA ({\bf k}))$ of the line bundle $\cO_\cA ({\bf k})$ can be represented by polynomials of multi-degree ${\bf k}$ in $R$, so we write
\begin{equation}
 H^0(X,\cO_\cA ({\bf k}))\cong R_{\bf k}\; . \label{H0cL}
\end{equation}
A co-dimension $K$ CICY manifold $X\subset\cA$ is defined as the zero
locus of homogeneous polynomials $p_1,\ldots ,p_K$ and we denote the
normal bundle of $X$ in $\cA$ by $\cN$. We define line bundles on $X$
by restricting ambient space line bundles, that is $\cO_X({\bf
  k})\equiv\cO_\cA({\bf k})|_X$. Moreover, we assume that the CICY
manifold is ``favourable'', that is, all line bundles on $X$ are
obtained in this way. The coordinate ring of $X$ is given by
\begin{equation}
 A=\frac{R}{\langle p_1,\ldots ,p_K\rangle}\; ,
\end{equation}
and it inherits the multi-grading from $R$.  We denote by $A_{\bf k}$ the multi-degree ${\bf k}$ part of $A$.

For the purpose of this appendix, we focus on line bundles
$\cL=\cO_\cA ({\bf k})$ and their counterparts $L=\cO_X({\bf k})$ on
$X$ which satisfy the vanishing conditions~\footnote{This can be
  slightly weakened without changing the result of this appendix. In
  fact we require the vanishing conditions stated in Table
  \ref{tablecond}.}
\begin{table}[t]\begin{center}
\begin{tabular}{cc}
  $H^{q-1}(\cA, \wedge^q {\cal N}^* \otimes \cL) =0 \qquad \forall \; 2 \leq q \leq K-1 \;,$& $H^{q-2}(\cA, \wedge^q {\cal N}^*\otimes \cL) =0 \qquad
  \forall \; 3 \leq q \leq K-1 \;,$ \\
  $H^q (\cA, \wedge^q {\cal N}^* \otimes \cL) = 0 \qquad
  \forall \; 2 \leq q \leq K-2 \;, $& 
  $ H^{q-1} (\cA, \wedge^q {\cal N}^* \otimes \cL) =0 \qquad
  \forall \; 3 \leq q \leq K-2\;,$ \\$ H^{K-1}(\cA,
  \wedge^{K} {\cal N}^*\otimes \cL) = 0 \;, $   & $
  H^{K-2} (\cA, \wedge^K {\cal N}^* \otimes \cL) =0 \;,$
\end{tabular}
\begin{tabular}{cc}
$H^q(\cA, \wedge^q {\cal N}^*\otimes \cL) =0 \qquad \forall
\; 1 \leq q \leq K-1 \;,$\\ $H^{q+1}(\cA,\wedge^q {\cal N}^*
\otimes \cL)=0 \qquad \forall \; 1 \leq q \leq K-2 \;,$\\$
H^K(\cA, \wedge^K {\cal N}^* \otimes \cL)=0 \;.$
\end{tabular}
\mycaption{The cohomology conditions that must be satisfied in
  order for the conclusions of Appendix A to hold.}\label{tablecond}
\end{center}
\end{table}
 \begin{equation}
 H^q(\wedge^\kappa\cN^\star\otimes\cL)=0 \label{vancond}
\end{equation}
for  $q>0$ and $\kappa = 0,\ldots ,K$. We note that, as a consequence of Kodaira's vanishing theorem applied to line bundles $\cL$ on the ambient space $\cA$, all positive line bundles fall into this class. Provided the above vanishing conditions are satisfied we want to show that
\begin{equation}
 H^0(X,L)\cong A_{\bf k}\; . \label{H0L}
\end{equation}

It is instructive to first do this for co-dimension one CICY manifolds, that is, for $K=1$, before embarking on the general case. For $K=1$ the Koszul resolution of $L$ is given by the short exact sequence
\begin{equation}
 0\rightarrow \cN^\star\otimes\cL\stackrel{\cdot p}{\longrightarrow}\cL\rightarrow L\rightarrow 0\; ,
\end{equation} 
where $\cdot p$ denotes multiplication by the defining polynomial $p$ of $X$. This leads to the long exact sequence
\begin{eqnarray}
 0&\rightarrow& H^0(\cA,\cN^\star\otimes\cL)\stackrel{p}{\longrightarrow} H^0(\cA,\cL)\rightarrow H^0(X,L)\nonumber\\
   &\rightarrow& H^1(\cA,\cN^\star\otimes\cL)\longrightarrow\dots\; .
 \end{eqnarray}  
Since $H^1(\cA,\cN^\star\otimes\cL)=0$ from the above vanishing assumptions one concludes that
\begin{equation}
 H^0(X,L)\cong\frac{H^0(\cA,\cL)}{p\left(H^0(\cA,\cN^\star\otimes\cL)\right)}\; .
\end{equation} 
Combining this with Eq.~\eqref{H0cL} the desired statement~\eqref{H0L} follows immediately.

We now proceed to the case of general co-dimension $K$. While the basic structure of the argument remains unchanged from the $K=1$ case a technical complication arises because the Koszul sequence
\begin{equation}
 0\rightarrow\wedge^K\cN^\star\otimes\cL\rightarrow\wedge^{K-1}\cN^\star\otimes\cL\rightarrow\dots\rightarrow\wedge^\kappa\cN^\star\otimes\cL\rightarrow\dots\rightarrow\cN^\star\otimes\cL\rightarrow\cL\rightarrow L\rightarrow 0
\end{equation}
is no longer short.  A simple way of dealing with this is to break the sequence up into short exact sequences
\begin{equation}
\begin{array}{lllllllll}
 0&\rightarrow&\wedge^K\cN^\star\otimes\cL&\rightarrow& \wedge^{K-1}\cN^\star\otimes\cL&\rightarrow& C_{K-1}&\rightarrow &0\\
&&&&\vdots&&&\\
0&\rightarrow& C_{\kappa +1}&\rightarrow&\wedge^\kappa\cN^\star\otimes\cL&\rightarrow& C_\kappa&\rightarrow& 0\label{klong}\\
&&&&\vdots&&&\\
0&\rightarrow& C_1&\rightarrow&\cL&\rightarrow& L&\rightarrow& 0\; ,
\end{array} 
\end{equation}
introducing co-kernels $C_1,\ldots ,C_{K-1}$. Here, we have $\kappa = 1,\ldots ,K-2$ in the middle sequence. From our vanishing condition, the first of these sequences implies that $H^q(\cA,C_{K-1})=0$ for all $q>0$. Further, from the long exacts sequence associated to the middle sequence above and the vanishing conditions it follows that $H^{q-1}(\cA,C_\kappa)\cong H^q(\cA,C_{\kappa +1})$ for $\kappa =1,\ldots ,K-2$ and $q=2,\ldots K+3$. Together, this means that $H^1(\cA ,C_\kappa )=0$ for $\kappa =1,\ldots ,K-1$ and, hence, the long exact sequences associated to \eqref{klong} all break after three terms. This leads to the recursion relations
\begin{eqnarray}
 H^0(X,L)&\cong&\frac{H^0(\cA,\cL)}{H^0(\cA,C_1)}\\
 H^0(\cA,C_\kappa)&\cong&\frac{H^0(\wedge^\kappa\cN^\star\otimes\cL)}{H^0(\cA,C_{\kappa+1})}\\
 H^0(\cA,C_{K-1})&\cong&\frac{H^0(\cA,\wedge^K\cN^\star\otimes\cL)}{H^0(\cA\wedge^{K-1}\cN^\star\otimes\cL)}\; ,
\end{eqnarray} 
where $\kappa = 1,\ldots ,K-2$, which allow one to express $H^0(X,L)$
as a ``chain of quotients''. However, since
\begin{equation}
0\rightarrow H^0(\cA,\wedge^K\cN^\star\otimes \cL)\rightarrow H^0(\cA,\wedge^{K-1}\cN^\star\otimes\cL)\rightarrow\dots\rightarrow H^0(\cA,\cN^\star\otimes L)\rightarrow H^0(\cA,\cL)\rightarrow H^0(X, L)\rightarrow 0 
\end{equation}
is a complex it is sufficient to keep the first quotient in this chain. Hence, we have
\begin{equation}
 H^0(X,L)\cong\frac{H^0(\cA,\cL)}{p\left(H^0(\cA,\cN^\star\otimes\cL)\right)}
\end{equation} 
where $p$ is the map induced by the defining polynomials $p_1,\ldots ,p_K$ of the CICY manifold. Using the polynomial representatives~\eqref{H0cL} for sections of line bundles in the ambient space this implies the desired Eq.~\eqref{H0L}.

\section{Proof of Equivalence of Formulations}\label{ap:equal}
\setall
In this Appendix, as promised in the text, we give a formal mathematical proof of why calculating the Yukawa couplings using \eqref{physyukawa} is equivalent to the  maps in cohomology as given in Section \ref{s:coho}.
\subsection{Chain complexes and bicomplexes}

The standard way to convert a bicomplex\footnote{Warning: the standard
 definition of {\em double complex} in \cite[p.~60]{MR0077480} (see also
 \cite[p.~174, Exer.~11]{MR2327161} and \cite[p.~8]{MR1269324}) uses a sign
 convention different from the one we use here, namely $d' d'' = - d'' d'$.}
$C = C_{\cdot\cdot}$ (with horizontal differential $d' : C_{p,q} \to C_{p-1,q}$
and vertical differential $d'' : C_{p,q} \to C_{p,q-1}$ that commute) to a
chain complex is to define the {\em total complex} $\Tot C$ by setting $(\Tot
C)_n = \bigoplus_{p+q=n} C_{p,q}$ and setting the differential $d : \Tot C \to
\Tot C$ of degree $-1$ to be $x \mapsto d'(x) + (-1)^p d''(x)$ for $x \in
C_{p,q}$.  This is in accordance with the principle of signs \cite{MR0209338}
if $C$ is the tensor product of two chain complexes and we identify the symbols
$d$, $d'$, and $d''$.

Given $m \in \IZ$ and a chain complex $B$ define the shifted chain complex
$B[m]$ by setting $B[m]_p = B_{m+p}$; use the same differential, with no change
in sign\footnote{Warning: this sign convention differs from the standard one
 implied by \cite[p.~72, Exercise 1]{MR0077480} and explicitly presented in
 \cite[p.~9]{MR1269324}.  There the differential on $B[m]$ is equal to
 $(-1)^m$ times the differential on $B$.  Better notation for that concept,
 compatible with the principle of signs \cite{MR0209338}, would be $[m]B$.  An
 isomorphism $[m]B \xrightarrow \isom B[m]$, also compatible with the principle
 of signs, can be defined by $x \mapsto (-1)^{mp} x$ for $x \in B_p$.}.
Similarly, given $m,n \in \IZ$ and a bicomplex $C$ define the shifted bicomplex
$C[m,n]$ by setting $C[m,n]_{p,q} = C_{m+p,n+q}$.

The formula for the differential in $\Tot C$ involves $p$ but not $q$, so there
is a simple isomorphism $\Tot(C[0,n]) \isom (\Tot C)[n]$ involving just direct
sums of identity maps, with no minus signs involved.  Thus, if we think of a
bicomplex $C$ as being assembled from its rows $C_{\cdot,q}$ for $q \in \IZ$,
reindexing the rows results in shifting the total complex.  If the bicomplex is
zero outside the range $0 \le q \le N$ we will use the pictorial notation $C =
[C_{\cdot,0} \from C_{\cdot,1} \from \dots \from C_{\cdot,N}]$ to indicate its
assembly from its rows.  No minus signs are to be used when assembling a
bicomplex from chain complexes and maps between them in this way.

In general, an isomorphism of chain complexes $\gamma : \Tot(C[m,n])
\xrightarrow \isom (\Tot(C))[m+n]$ can be defined as $(-1)^{mq}$ times the
identity map on the component $(C[m,n])_{p,q} = C_{p+m,q+n}$.  We omit the
computation that $\gamma$ is a chain map.  A careful eye can discern something
of degree $m$ moving past something of degree $q$, in accordance with the
principle of signs \cite{MR0209338}.

Given a map $f : B \to C$ of chain complexes we define the mapping
cone\footnote{Our definition of the mapping cone is not the usual one, see
 \cite[p.~18,~20]{MR1269324}.} by setting $\Cone f = \Cone(C \from B) = \Tot
[C \from B]$.  There are isomorphisms $\Cone(C \from 0) \isom C$ and $\Cone(0
\from B) \isom B[-1]$, and the exact sequence $0 \to [C \from 0] \to [C \from
 B] \to [0 \from B] \to 0$ of bicomplexes leads to an exact sequence $0 \to C
\to \Cone f \to B[-1] \to 0$ of chain complexes.  Given $c \in C_p$ and $b \in
B_{p-1}$ the element $(c,b) \in (\Cone f)_p$ satisfies $d(c,b) =
(dc+(-1)^{p-1}fb,db)$.

A fundamental lemma in homological algebra states that a map $C \to D$ of first
quadrant bicomplexes that is a quasi-isomorphism in each row induces a
quasi-isomorphism on total complexes.  The same statement applies to a map of
third quadrant bicomplexes, or when rows are replaced by columns.  A slightly
stronger version, for filtered complexes, is proved in \cite[Lemma
 3.2]{MR1949551}.  This result is presented as the {\em acyclic assembly
 lemma} in \cite{MR1269324}.

Given a short exact sequence $E : 0 \to A \xrightarrow f B \xrightarrow g C \to
0$ of chain complexes, the corresponding map $[0 \from A] \to [C \from B]$ of
bicomplexes is a quasi-isomorphism in each column, hence, according to the
lemma, $A[-1] \to \Cone(C \from B)$ is a quasi-isomorphism.  Its inverse in the
derived category composed with the map $C \to \Cone(C \from B)$ gives a map
$\rho = \rho_E : C \to A[-1]$ in the derived category.  We would like to
compare the induced map $\rho : H_pC \to H_{p-1}A$ with the connecting
homomorphism $\partial = \partial_E : H_pC \to H_{p-1}A$ that appears in the long
exact homology sequence.  Given cycles $c \in C_p$ and $a \in A_{p-1}$,
$\partial[c] = [a]$ means that there is an element $b \in B_p$ such that $gb=c$
and $fa=db$.  The element $(0,b) \in \Cone(C \from B)_{p+1}$ satisfies $d(0,b)
= ((-1)^{p}gb,db) = ((-1)^{p}c,fa)$, so $((-1)^{p-1}c,0)$ and $(0,fa)$ are
homologous elements of $\Cone(C \from B)_{p}$, which tells us that $\rho[c] =
(-1)^{p-1}[a]$, and thus $\rho = (-1)^{p-1} \partial$ on $H_pC$.

Now we consider longer extensions in the sense of Yoneda.  Suppose we have an
exact sequence $E : 0 \to A \to B_n \to \dots \to B_2 \to B_1 \to C \to 0$ of
chain complexes.  The commutative diagram
\[
\xymatrix{
 \cdots & 0 \ar[d]\ar[l] & 0 \ar[d]\ar[l] & 0 \ar[d]\ar[l] & \cdots \ar[l] & 0 \ar[d]\ar[l] & A \ar[d]\ar[l] & 0 \ar[d]\ar[l] & \cdots \ar[l] \\
 \cdots & 0 \ar[l] & C \ar[l] & B_1 \ar[l] & \cdots \ar[l] & B_{n-1} \ar[l] & B_n \ar[l] & 0 \ar[l] & \cdots \ar[l] \\
}\]
of chain complexes can be regarded as a map of chain complexes of chain
complexes.  The corresponding map $[0 \from \dots \from A] \to [C \from B_1
 \from \dots \from B_n]$ of bicomplexes is a quasi-isomorphism in each column
(by exactness of E), hence the map $A[-n] \to \Tot[C \from B_1 \from \dots
 \from B_n]$ is a quasi-isomorphism.  Its inverse composed with the map $C \to
\Tot[C \from B_1 \from \dots \from B_n]$ gives a map $\rho = \rho_E : C \to
A[-n]$ in the derived category.

Suppose we have another exact sequence $F : 0 \to C \to P_m \to \dots \to P_2
\to P_1 \to Q \to 0$ of chain complexes, and consider the associated map
$\rho_F : Q \to C[-m]$.  Let $E*F : 0 \to A \to B_n \to \dots \to B_2 \to B_1
\to P_m \to \dots \to P_2 \to P_1 \to Q \to 0$ be the exact sequence obtained
by splicing $E$ to $F$ along $C$; the differential in the middle is the
composite map $B_1 \to C \to P_m$.  The following commutative diagram of
bicomplexes, in which quasi-isomorphisms are indicated by $\sim$, shows that
$\rho _ {E*F} = \rho_E[-m] \circ \rho_F$; the simplicity of this formula and
the absence of signs in it is a consequence of our choices above.
\[
\xymatrix{
 & 
 & A[-m-n] \ar[d]^\sim \\
 & C[-m] \ar[d]^\sim \ar[r]
 & [C \from B_1 \from \dots \from B_n][-m] \ar[d]^\sim \\
   Q \ar[r]
 & [Q \from P_1 \from \dots \from P_m] \ar[r]
 & [Q \from P_1 \from \dots \from P_m \from B_1 \from \dots \from B_n]
\\}\]

Now let's decompose our original sequence $E$ by writing it as $E = E_1 * \dots
* E_n$, where $E_i : 0 \to D_i \to B_i \to D_{i-1} \to 0$, and $D_n = A$, $D_0
= C$, and $D_i = \im(B_{i+1} \to B_i)$ for $0 < i < n$.  Then $\rho_E =
\rho_{E_1}[-(n-1)] \circ \dots \circ \rho_{E_{n-1}}[-1] \circ \rho_{E_n}$.  The
resulting map $H_p(\rho_E) : H_pC \to H_{p-n}A$ is thus a composite of
connecting homomorphisms, up to sign.  More precisely, $H_p(\rho_E) =
H_p(\rho_{E_1}[-(n-1)] \circ \dots \circ \rho_{E_{n-1}}[-1] \circ \rho_{E_n}) =
H_p(\rho_{E_1}[-(n-1)]) \circ \dots \circ H_p(\rho_{E_{n-1}}[-1]) \circ
H_p(\rho_{E_n}) = H_{p-n+1}(\rho_{E_1}) \circ \dots \circ
H_{p-1}(\rho_{E_{n-1}}) \circ H_{p}(\rho_{E_n}) = ((-1)_{p-n}\partial_{E_1})
\circ \dots \circ ((-1)_{p-2}\partial_{E_{n-1}}) \circ
((-1)^{p-1}\partial_{E_n}) = (-1)^{(p-n)+\dots+(p-2)+(p-1)} \partial_{E_1}
\circ \dots \circ \partial_{E_{n-1}} \circ \partial_{E_n} = (-1)^{pn+n(n+1)/2}
\partial_{E_1} \circ \dots \circ \partial_{E_{n-1}} \circ \partial_{E_n}$.
(This result was proved in \cite[Chap.~V, Prop.~7.1, p.~92]{MR0077480}.  See
also the application in \cite[Chap.~V, Exer.~8, p.~105]{MR0077480}.)  In
particular, $H_0(\rho_E) = (-1)^{n(n+1)/2} \partial_{E_1} \circ \dots \circ
\partial_{E_{n-1}} \circ \partial_{E_n}$.

Now suppose our chain complexes are bounded above and have their
components drawn from an abelian category $\cC$ with enough
injectives, and suppose we are studying the right derived functors
$R^pF$ of a left exact additive functor $F : \cC \to \cV$, where $\cV$
is an abelian category.  A chain complex $B$ with $B_p$ injective for
each $p$ is called {\em injective}.  If $E : \dots \to C_2 \to C_1 \to
C_0 \to \dots$ is a chain complex of such complexes (each bounded
above), then it maps (injectively) to a chain complex $E' : \dots \to
C'_2 \to C'_1 \to C'_0 \to \dots$ of injective chain complexes (each
bound above), so that for each $p$ the map $C_p \to C'_p$ is a
quasi-isomorphism; moreover, if $E$ is exact, then $E'$ may be chosen
to be exact; also, $E'$ may be chosen so that $C'_p = 0$ for all $p$
with $C_p = 0$.\footnote{We only sketch the proof. As in the
  construction of a Cartan-Eilenberg resolution of a chain complex,
  one writes the chain complex in terms of short exact sequences with
  maps from the tail end of one to the start of the next.  Then one
  modifies the proof that the modules in a short exact sequence have
  injective resolutions that fit into a short exact sequence by
  replacing cokernels by pushouts at a certain point.  In any case,
  for our intended application to sheaves on a topological space, we
  don't really need this abstract formulation, because of the
  canonical Godement flasque resolution.}

In particular, a chain complex $C$ maps via an injective quasi-isomorphism to
an injective chain complex $C'$ (an injective resolution).  We set $RF(C) =
F(C')$ and $R^pF(C) = H^p(F(C'))$, thereby extending the usual definition of
$RF$ for objects {\em (cohomology)} of our category to chain complexes {\em
 (hypercohomology)}.  The usual arguments that show this definition is
independent of the choice of injective resolution can be extended to cover this
case.  See \cite[Chap.~XVII]{MR0077480} for a detailed discussion of
hyperhomology and hypercohomology.  See also \cite[p.~183, Exer.~17]{MR2327161}
for a discussion of resolutions of complexes.

When working with chain complexes of sheaves of abelian groups on a space $X$,
we will use the same notation for sheaf cohomology and for sheaf
hypercohomology, writing $H^p(X,C)$ whether $C$ is a sheaf or a complex of
sheaves (bounded above).  In this context, one may use the flasque resolution
$C \to G(C)$ constructed by Godement in \cite[Chap.~II, Sect.~4.3,
 p.~167]{MR0102797}.  The construction gives an exact functor from sheaves to
flasque resolutions of them, hence, by applying it to each sheaf in a chain
complex and then taking the total complex, it gives an exact functor from chain
complexes to injective resolutions of them.

Whether we use injective resolutions or flasque resolutions, the formula above
for $H_0(\rho_E)$ leads to an analogous formula on sheaf cohomology for
$H^0(X,\rho_E) : H^0(X,C) \to H^n(X,A)$ as a composite of connecting
homomorphisms with a sign.

\subsection{Cup products in hypercohomology of sheaves}

In this section, the tensor product $B \otimes C$ of sheaves $B$ and $C$ on $X$
may denote either: tensor product of sheaves of abelian groups; tensor product
of sheaves of $R$-modules, where $R$ is a commutative ring; or tensor product
of sheaves of $\cO$-Modules, where $\cO$ is a sheaf of rings on $X$.  It is an
additive functor in each variable.

Godement's canonical flasque resolution $G(C)$ of a sheaf $C$ in
\cite[Chap.~II, Sect.~4.3, p.~167]{MR0102797} begins with the map $\eta : C \to
G^0(C) = \prod_{x \in X} (i_x)_* C_x$, where $i_x : \{x\} \to X$ is the
inclusion map.  The stalk of $\eta$ at any point $y \in X$ can be split by
projecting onto the factor corresponding to $y$ in the product, and that shows
$\eta$ is injective.  Moreover, if $B$ is another sheaf, then $B \otimes \eta$
is an injective map, for the same reason.  The second step in Godement's
construction is $G^1(C) = G^0(\coker \eta)$, and the pattern continues.
Because tensor product is always right exact, it follows that $B \otimes C \to
B \otimes G(C)$ is a quasi-isomorphism, and that is also true when $C$ is a
chain complex of sheaves.  This fact, which we may call {\em (universal
 exactness)}, makes cup product operations possible, as we shall now see.
(Our approach differs slightly from Godement's original approach to cup
products in \cite[Chap.~II, Sect.~6.1, p.~238]{MR0102797}, in that he
emphasized the role of external tensor product sheaves on $X \times X$.  For a
thorough and modern approach, see \cite{MR1731435}.)

If $B$ and $C$ are chain complexes, the bicomplex $B \otimes C$ is defined by
setting $(B \otimes C)_{p,q} = B_p \otimes C_q$.  The vertical and horizontal
differentials come from those of $B$ and $C$, with no signs introduced,
contrary to the standard convention \cite[Chap.~4, Sect.~5]{MR0077480}.

By universal exactness of the Godement resolution, the map $\Tot(B \otimes C)
\to \Tot(G(B) \otimes G(C))$ is a quasi-isomorphism, hence the identity map on
$\Tot(B \otimes C)$ can be lifted to a map from $\Tot(G(B) \otimes G(C))$ to an
injective resolution of $\Tot(B \otimes C)$, unique up to homotopy.  The
resulting pairing $H^p(X,B) \otimes H^q(X,C) \to H^{p+q}(X,\Tot(B \otimes C))$
is the {\em cup product} in hypercohomology.  If a map $\Tot(B \otimes C) \to
D$ is given, the resulting composite pairing $H^p(X,B) \otimes H^q(X,C) \to
H^{p+q}(X,D)$ may also be called a cup product pairing.  We may also assemble
these maps into a single map $H^*(X,B) \otimes H^*(X,C) \to H^*(X,D)$ of graded
groups.

Let's examine compatibility of cup products with shifting.  Composition with
the map $\gamma$ introduced above gives a map $\Tot(B[m] \otimes C[n]) \to
D[m+n]$ that leads to a cup product pairing $H^p(X,B[m]) \otimes H^q(X,C[n])
\to H^{p+q}(X,D[m+n])$.  Identity maps can be used to compare this with the
original cup product pairing $H^{p-m}(X,B) \otimes H^{q-n}(X,C) \to
H^{p+q-m-n}(X,D)$, and the resulting discrepancy is the factor $(-1)^{mq}$
appearing in the definition of $\gamma$.

\subsection{Symmetric and exterior powers of complexes}

Suppose $k \ge 0$ and $C$ is a chain complex.  Let $C^{\otimes k}$ denote the
tensor product $C \otimes \dots \otimes C$ of $k$ copies of $C$.  The symmetric
group $\Sigma_k$ acts on $C^{\otimes k}$ by permuting the factors, but a sign
must be inserted to get an action on $\Tot C^{\otimes k}$, in accordance with
the principle of signs \cite{MR0209338}.  Transposing adjacent factors involves
a minus sign exactly when the two factors are both of odd degree, and the total
sign can be determined by writing an arbitrary permutation as a product of
adjacent transpositions.  Another way of saying it that one excises the factors
of even degree, collapsing to a possibly shorter tensor product, and then takes
the sign of the residual permutation on the factors of odd degree.

To see that that works, it suffices to consider the case $k=2$.  Let $\tau :
C_p \otimes C_q \to C_q \otimes C_p$ denote the (signed) transposition map
defined by $x \otimes y \mapsto (-1)^{pq} y \otimes x$.  If $x \in C_p$ and $y
\in C_q$, then $\tau(d(x \otimes y)) = \tau(dx \otimes y + (-1)^p x \otimes dy)
= (-1)^{(p+1)q } y \otimes dx + (-1)^{p+p(q+1)} dy \otimes x = (-1)^{pq} dy
\otimes x + (-1)^{pq+q} y \otimes dx = d((-1)^{pq} y \otimes x) = d(\tau(x
\otimes y)) $.

Assume for the rest of the section that we are working with coherent sheaves on
a variety $X$ over a field of characteristic $0$, and that the tensor product
$B \otimes C$ of sheaves denotes tensor product of sheaves of $\cO_X$-Modules.
Assume that $B$ is a locally free finitely generated $\cO$-Module (vector
bundle).  Then $B \otimes C$ is an exact functor of the sheaf $C$.  Hence, if
$C$ is an acyclic chain complex, so is $B \otimes C$.  If $C \to D$ is a
quasi-isomorphism, then so is $B \otimes C \to B \otimes D$ (because being a
quasi-isomorphism is determined by whether the mapping cone is acyclic, and
formation of the mapping cone commutes with tensor product by $B$).
Alternatively, assume that $B$ is a chain complex of locally free sheaves, and
that all our chain complexes are bounded above.  Then if $C \to D$ is a
quasi-isomorphism, then so is $\Tot(B \otimes C) \to \Tot(B \otimes D)$.
Finally, if $C \to D$ is a quasi-isomorphism of chain complexes of locally free
sheaves, then $\Tot C^{\otimes k} \to \Tot D^{\otimes k}$ is a
quasi-isomorphism.

Now let $C$ be a chain complex, let $S^k C$ denote the part of $\Tot C^{\otimes
 k}$ upon which $\Sigma_k$ acts trivially, and let $\wedge^k C$ denote the
part of $\Tot C^{\otimes k}$ upon which $\Sigma_k$ acts by the sign of
permutations.  The projection operators $(1/k!) \sum_{\sigma \in \Sigma_k}
\sigma$ and $(1/k!)  \sum_{\sigma \in \Sigma_k} (-1)^\sigma \sigma$ show that
$S^k C$ and $\wedge^k C$ appear functorially as direct summands of $\Tot
C^{\otimes k}$.  Moreover, if $C \to D$ is a quasi-isomorphism, then so are the
induced maps $S^k C \to S^k D$ and $\wedge^k C \to \wedge^k D$.

Let's compute symmetric and exterior powers of complexes of length $0$ and
length $1$ in terms of symmetric and exterior powers of sheaves.  Suppose $C$
is a sheaf.  When suggested by the notation, we convert $C$ to a chain complex
of length $0$ by putting it in degree $0$ and putting zeroes in the other
positions; thus $C[m]$ will denote the chain complex of length $0$ with $C$ in
position $-m$ and zeroes in the other positions.  With this notation, we see
that $S^k(C[m]) = (S^kC)[km]$ if $m$ is even and $S^k(C[m]) = (\wedge^k C)[km]$
if $m$ is odd, and $\wedge^k(C[m]) = (\wedge^kC)[km]$ if $m$ is even and
$\wedge^k(C[m]) = (S^k C)[km]$ if $m$ is odd.  Suppose now that $C = [A
 \xleftarrow d B]$ is a complex of length $1$; recall that this notation puts
$A$ in degree $0$ and $B$ in degree $1$.  We wish to compute $(S^kC)_q$ and
$(\wedge^kC)_q$ for $q \in \IZ$; for this purpose we may assume $d=0$, so that
$C \isom A \oplus B[-1]$.  Then $S^k C \isom S^k(A \oplus B[-1]) \isom
\bigoplus_{p+q=k} S^p A \otimes S^q(B[-1]) \isom \bigoplus_{p+q=k} S^p A
\otimes (\wedge^q B)[-q] \isom \bigoplus_{p+q=k} (S^p A \otimes \wedge^q
B)[-q]$.  The general conclusion is that $(S^k C)_q = S^{k-q} A \otimes
\wedge^q B$, and a similar argument shows that $(\wedge^k C)_q = \wedge^{k-q} A
\otimes S^q B$.

Suppose $E : 0 \to V \to B\to C \to 0$ is a short exact sequence of vector
bundles.  Then the map $V \to [C \from B][1]$ is a quasi-isomorphism, and
hence so are the maps $S^k V \to S^k([C \from B][1])$ and $\wedge^k V \to
\wedge^k([C \from B][1])$.  In other words, we have long exact sequences
$$ S^k E : 0 \to S^k V \to S^k B \to S^{k-1}B \otimes C \to \dots \to S^{k-p} B \otimes
\wedge^p C \to \dots \to \wedge^k C \to 0$$ and
$$ \wedge^k E : 0 \to \wedge^k V \to \wedge^k B \to \wedge^{k-1}B \otimes C \to \dots \to
\wedge^{k-p} B \otimes S^p C \to \dots \to S^k C\to 0.$$

(Suppose alternatively that $0 \to A \to B \to V \to 0$ is an exact sequence of
vector bundles.  Then we have long exact sequences
$$ 0 \to S^k A \to S^{k-1}A \otimes B \to \dots \to S^{k-p} A \otimes
\wedge^p B \to \dots \to \wedge^k B \to \wedge^k V \to 0$$ and
$$ 0 \to \wedge^k A \to \wedge^{k-1}A \otimes B \to \dots \to \wedge^{k-p} A
\otimes S^p B \to \dots \to S^k B \to S^k V \to 0.)$$

For an alternative presentation of the portion of these results that hold in
any characteristic, see \cite[Sect.~2]{MR1187703}.  For the case where $V=0$
see the exposition in \cite[p.~151, Ex.~1]{MR2327161}.

Our goal now is to relate $\rho_E$ to $\rho_{S^kE}$.  We have a commutative
diagram
\[
\xymatrix{
 (H^1 (X,V))^{\otimes k} \ar[r] \ar[d]^\isom &
    H^k(X, S^k V) \ar[d]^\isom \\
 (H^1 (X,[C \from B][1]))^{\otimes k} \ar[r]&
    H^k (X, S^k ([C \from B][1])) \\
 (H^1 (X,C[1]))^{\otimes k} \ar[r] \ar[u] &
    H^k (X, S^k (C[1])) \ar[u],
 }
\]
where the horizontal maps are obtained by iterating the cup product pairings.
Let $E^k = S^k$ if $k$ is even, and $E^k = \wedge^k$ if $k$ is odd.  Shifting
the bottom row of the diagram above yields the following diagram.  The vertical
maps arise from identity maps, and the diagram commutes up to sign of
$(-1)^{((k-1)+(k-2)+\dots+1)} = (-1)^{k(k-1)/2} $, using our earlier
computation.
\[
\xymatrix{
 (H^1 (X,C[1]))^{\otimes k} \ar[r] &
    H^k (X, S^k (C[1]))
\\
 (H^0 (X,C))^{\otimes k} \ar[r]\ar[u]^\isom &
    H^0 (X,E^k C) \ar[u]^\isom
 }
\]
Splicing the two diagrams together and retaining only the top and bottom rows
yields the following diagram, which commutes up to a sign of $(-1)^{k(k-1)/2}$.
\[
\xymatrix{
 (H^1 (X,V))^{\otimes k} \ar[r] &
    H^k(X, S^k V) 
\\
 (H^0 (X,C))^{\otimes k} \ar[r]\ar[u]_{(\rho_E)^{\otimes k}} &
    H^0 (X,E^k C) \ar[u]_{\rho_{S^k E}}
 }
\]
The vertical maps can also be expressed in terms of connecting homomorphisms,
as we have seen before.

\newpage


\end{document}